\documentclass{emulateapj}
\usepackage{graphicx,hyperref,amsmath}
\hypersetup{colorlinks,citecolor=black,filecolor=black,linkcolor=green,urlcolor=magenta,pdftex}
\usepackage{epstopdf}
\usepackage{epsfig}

\newcommand{\Fig}[1]{Fig.~\ref{#1}}

\newcommand{\Eqn}[1]{Eq.~\ref{#1}}
\newcommand{\lya}{Ly$\,\alpha$}
\newcommand{\lyb}{Ly$\, \beta$}
\newcommand{\lyc}{Ly$\, \gamma$}

\newcommand{\pow}[2]{\ensuremath{#1 \times 10^{#2}}}

\newcommand{\mpch}{\ensuremath{\,{\rm Mpc}/h\,}}

\newcommand{\avg}[1]{\ensuremath{\langle #1 \rangle}}
\newcommand{\bma}{\begin{math}}
\newcommand{\ema}{\end{math}}
\newcommand{\beq}{\begin{equation}}
\newcommand{\eeq}{\end{equation}}
\newcommand{\beqa}{\begin{eqnarray}}
\newcommand{\eeqa}{\end{eqnarray}}
\newcommand{\bc}{\begin{center}}
\newcommand{\ec}{\end{center}} 
\newcommand{\bit}{\begin{itemize}}
\newcommand{\eit}{\end{itemize}}
\newcommand{\kms}{{\rm km}\ensuremath{/}{\rm s}}
\newcommand{\cm}{\text{cm}}

\newcommand{\taudwa}{\tau_{\text{Ly}\alpha}^{\text{DW}}}
\newcommand{\taudwb}{\tau_{\text{Ly}\beta}^{\text{DW}}}
\newcommand{\vext}{v_{\text{ext}}}

\font\BFd=cmmib10
\font\BFt=cmmib10
\font\BFs=cmmib10 scaled 700
\font\BFss=cmmib10 scaled 500

\def\bbox#1{%
\relax\ifmmode
\mathchoice
{{\hbox{\BFd #1}}}
{{\hbox{\BFt #1}}}
{{\hbox{\BFs #1}}}
{{\hbox{\BFss #1}}}
\else \mbox{#1} \fi }

\newcommand{\xhi}{x_{\text{HI}}}
\newcommand{\axhi}{\langle x_{\text{HI}}\rangle}
\newcommand{\dd}{\text{d}}

\begin{document}
 
\submitted{\today. To be submitted to \apj.} 

\title{How to Search for Islands of Neutral Hydrogen in the $z \sim 5.5$ IGM}
\author{Matthew Malloy\altaffilmark{1} \& Adam Lidz\altaffilmark{1}}
\altaffiltext{1} {Department of Physics \& Astronomy, University of Pennsylvania, 209 South 33rd Street, Philadelphia, PA 19104, USA}
\email{mattma@sas.upenn.edu}

\begin{abstract}
Observations of the Lyman-alpha (Ly-$\alpha$) forest may allow
reionization to complete as late as $z \sim 5.5$, provided the ionization
state of the intergalactic medium (IGM) is sufficiently inhomogeneous
at these redshifts.  In this case, significantly neutral islands
may remain amongst highly ionized gas with the ionized regions allowing
some transmission through the Ly-$\alpha$ forest.
This possibility has the important virtue that it is eminently
testable with existing Ly-$\alpha$ forest data. In particular, we describe
three observable signatures of significantly neutral gas in the $z \sim 5.5$
IGM. We use mock quasar spectra produced from numerical simulations of
reionization to develop these tests.
First, we quantify how the abundance and length of absorbed regions
in the forest increase with the volume-averaged neutral fraction in our
reionization model.
Second, we consider stacking the transmission profile around highly absorbed
regions in the forest. If and only if there is significantly neutral gas in the
IGM, absorption
in the damping wing of the Ly-$\alpha$ line will cause the transmission to
recover slowly as one moves from absorbed to transmitted portions of the
spectrum. Third, the deuterium Ly-$\beta$ line should imprint a small but
distinctive absorption
feature slightly blueward of absorbed neutral regions in the Ly-$\beta$ forest.
 We show that these tests can be carried out with existing Keck HIRES
spectra at $z \sim 5.5$, with the damping wing being observable for $\axhi \gtrsim 0.05$ and the deuterium feature observable with additional high-resolution spectra for $\axhi \gtrsim 0.2$.
\end{abstract}

\keywords{cosmology: theory -- intergalactic medium -- large scale
structure of universe}

\section{Introduction} \label{sec:intro}

It has been nearly half a century since \citet{1965ApJ...142.1633G} pointed out that the lack of prominent
absorption troughs, blueward of the Ly-$\alpha$ line in quasar spectra, implies that intergalactic
hydrogen is highly ionized. 
Only in the year 2001 were 
complete ``Gunn-Peterson'' absorption troughs finally revealed in 
the Ly-$\alpha$ forest of high redshift ($z \gtrsim 6$) quasars discovered using the Sloan Digital Sky Survey (SDSS)
\citep{Fan:2001ff,Becker:2001ee,Djorgovski:2001ez}. 
Although these prominent absorption troughs were discovered more than a decade ago, the precise interpretation
of the observations, and their implications for the reionization history of the universe,
remain unclear. One difficulty here relates to the large optical depth to
Ly-$\alpha$ absorption: near $z \sim 6$, the optical depth 
is $\tau_\alpha \sim 4 \times 10^5$ in a fully neutral IGM at the cosmic mean density \citep{1965ApJ...142.1633G}. Based on this, it is common to infer 
that the IGM must be highly ionized below $z \lesssim 6$,
at which point quasar spectra do show some transmission through the Ly-$\alpha$ line. In addition, it is clearly hard
to discern whether the gas above $z \gtrsim 6$ -- that does show complete absorption in the Ly-$\alpha$ line --
is mostly neutral or is only neutral at the level of about one part in ten-thousand or so (e.g. \citealt{Fan:2005es}); in either case,
the Ly-$\alpha$ line will be completely absorbed.

However, if reionization is sufficiently inhomogeneous and ends late, there may be some transmission through
the Ly-$\alpha$ forest {\em before reionization completes} \citep{Mesinger:2009mv,Lidz:2007mz}. Theoretical models of reionization show that
the IGM during reionization resembles a two-phase medium, containing a mixture of highly ionized
bubbles along with mostly neutral regions. The ionized bubbles grow and merge, eventually filling essentially the entire
volume of the IGM with ionized gas; the redshift at which this process completes is highly uncertain and still awaits definitive empirical constraint. In principle, the ionized bubbles may allow transmission through the
Ly-$\alpha$ forest even when some of the IGM volume is still in fact filled by neutral regions, i.e., before reionization completes.
This calls into question the conventional wisdom described above -- that the presence of transmission through the $z \lesssim 6$ forest
necessarily implies reionization completed by $z=6$ \citep{Lidz:2007mz,Mesinger:2009mv} -- strictly speaking, this conclusion follows only in 
the unrealistic case of a homogeneously-ionized
IGM. 

Indeed, some portions of the $z \sim 5-6$ Ly-$\alpha$ forest are completely absorbed, while other portions of the forest at these redshifts show
transmission through the Ly-$\alpha$ line. 
Quantitatively, if one counts only the fraction of pixels with some transmission
through the forest as ``certain to be ionized'', the volume-averaged neutral hydrogen fraction need only be smaller than 
$\avg{x_{\rm HI}} < 0.2$ at $5 \leq z \leq 5.5$, and smaller than $\avg{x_{\rm HI}} < 0.5$ at $z=6$ \citep{McGreer:2011dm}.
These constraints are conservative since even mostly-ionized gas will give rise to some completely absorbed regions at these redshifts, 
but it is nevertheless interesting to ask whether some of the absorbed regions could in fact
come from remaining ``islands'' of mostly neutral hydrogen gas in the IGM. The dark pixel fraction constraints of \citet{McGreer:2011dm} certainly leave
plenty of parameter space open for reionization completing at $z \leq 6$. 

In fact, there are hints -- albeit indirect ones -- that significant
amounts of neutral gas may remain in the IGM at these late times and so 
we believe that investigating this possibility amounts to {\em more} than closing a remaining ``loophole'' in the analysis
of the $z \lesssim 6$ Ly-$\alpha$ forest. For example, recent measurements of the rest-frame ultraviolet galaxy luminosity function 
suggest a relatively low ionizing emissivity at $z \gtrsim 5-6$, even for seemingly generous assumptions
about the escape fraction of ionizing photons ($f_{\rm esc} \sim 0.2$) and allowing significant extrapolations down the faint end of the luminosity function;  e.g. the preferred model
of \citet{Robertson:2013bq} (that matches these observations) has $\axhi = 0.1$ at $z=6$. In addition, the fraction of Lyman-break galaxies with detectable Ly-$\alpha$ emission lines shows evidence
for a rapid drop between $z \sim 6-7$ which may require a significant neutral fraction at $z \sim 7$ (e.g., \citealt{2012ApJ...744..179S,Pentericci:2014nia}, although see \citealt{2013MNRAS.429.1695B,Taylor:2013qia}). The inferred $z \sim 7$ neutral fraction here would be easier to accommodate if there is
still some neutral gas at $z \leq 6$.  Furthermore, \citet{Becker:2014oga} recently discovered an impressive $\sim 110$ Mpc/$h$ dark region in the $z \sim 5.7$ Ly-$\alpha$ forest. This may result from an upward opacity fluctuation -- driven by a fluctuating ultraviolet radiation field in a mostly ionized IGM -- but this striking observation invites contemplating the more radical possibility that diffuse neutral regions remain in the IGM at this late time.
Finally, \citet{Mesinger:2006kn} and \citet{2013MNRAS.428.3058S} argue that the proximity zones of quasars at $z \geq 6$ show evidence for damping wing absorption and a significant neutral fraction, further motivating the search for neutral gas at slightly later times.

Perhaps more importantly, we can design robust observational tests for the presence of neutral islands in the $z \sim 5.5$ IGM, and either
definitively detect neutral hydrogen at these redshifts, or significantly improve on the existing upper limits from \citet{McGreer:2011dm}. 
Towards this end, we study three possible tests for identifying neutral islands in the $z \sim 5-6$ IGM, each of which 
can be applied using existing 
Ly-$\alpha$ forest spectra. The presence of some transmission through the Ly-$\alpha$ forest at $z \leq 6$ allows us to consider
tests that can not be applied at still higher redshift where the forest is 
completely absorbed (asides for in the ``proximity zones'' close to the quasar itself). 
We develop these tests using mock quasar spectra extracted from the numerical reionization simulations of
\citet{McQuinn:2007dy}.
The first test we consider has been studied before (e.g. \citealt{Fan:2005es}, \citealt{2010MNRAS.407.1328M}, \citealt{McGreer:2011dm}), but is the most model 
dependent: the abundance and size distribution of ``dark gaps'', i.e., regions of saturated absorption in the Ly-$\alpha$ forest. Here we focus on
the plausible impact of inhomogeneous reionization on the dark gap statistics. The second test utilizes the fact that the natural line width of the Ly-$\alpha$ line gives rise to extended damping wing absorption, in
the case that highly neutral gas is present in the IGM \citep{MiraldaEscude:1997qb}. As a result, the transmission recovers more
slowly around significantly neutral absorbed regions than around absorbed yet ionized regions. We find that this
signature can be detected in partly neutral models by examining the stacked profile around extended absorbed regions. Note that, in contrast to previous work, here we propose to search
for the damping wing signature in typical regions of the IGM, as opposed to in the proximity zones of quasars (\citealt{Mesinger:2006kn,2013MNRAS.428.3058S}), or redward of Ly-$\alpha$ at the source redshift. 
Our third test involves the stacked profile
of extended absorbed regions in the Ly-$\beta$ forest. If these regions are significantly neutral, there should be a feature
from absorption in the deuterium Ly-$\beta$ line just blueward (but not redward) of absorbed regions.

The outline of this paper is as follows. In \S\ref{sec:Viability}, we briefly discuss which range of (volume-averaged) neutral fractions are physically plausible at $z=5.5$. In 
\S \ref{sec:Sims} we describe the simulations used and the process for generating mock spectra. We discuss how the dark gap size distribution may be used to constrain the neutral fraction in \S \ref{sec:HIDistributions}. In \S \ref{sec:Stacking}, we describe how quasar spectra may be stacked in order to reveal the presence of deuterium and HI damping wing absorption in an idealized scenario and discuss adapting this approach for more realistic spectra in \S \ref{sec:RealSpectra}. We then apply this approach to mock quasar spectra in \S \ref{sec:Results}, discuss the constraining power of the stacking approaches in \S \ref{sec:Forecasts}, and conclude in \S \ref{sec:Conclusion}. Throughout, we consider a $\Lambda$CDM cosmology parametrized by $n_{s} = 1$, $\sigma_{8} = 0.8$, $\Omega_{m} = 0.27$, $\Omega_{\Lambda} = 0.73$, $\Omega_{b} = 0.046$, and $h = 0.7$, (all symbols have their usual meanings), broadly consistent with recent Planck constraints from \cite{Ade:2013zuv}.

\section{Viability of Transmission Through a Partially Neutral IGM} \label{sec:Viability}

 Ideally, this study would make use of mock Ly-$\alpha$ forest spectra extracted from fully self-consistent simulations of reionization, in which the efficiency of the ionizing sources and other relevant parameters are tuned so that reionization completes at $z \leq 6$.
 Unfortunately, large-scale reionization simulations that simultaneously resolve the properties of the gas distribution, as well as the sources and sinks of ionizing photons, while
 capturing large enough volumes to include a representative sample of the ionized regions, are still quite challenging. Here, we instead explore more approximate, yet more flexible, models.
 As we describe in more detail in the next section, we make use of the reionization simulations of \citet{McQuinn:2007dy} to describe the size  and spatial distribution of the ionized and
 neutral regions during reionization. Inside of the ionized regions, we rescale the simulated photoionization rates, adjusting the intensity of the UV radiation field to match the observed mean transmitted flux through the Ly-$\alpha$ forest.
 For simplicity, we assume  that the intensity of the UV radiation field in the ionized regions is uniform and comment on the possible impact of this approximation where relevant.
 
 Before proceeding further, however, it is worth considering which (volume-averaged) neutral fractions are physically plausible near $z \sim 5.5$. In order to get transmission through
 the $z \sim 5.5$ Ly-$\alpha$ forest, at least some of the hydrogen needs to be highly ionized.  This requires the mean free path of ionizing photons to be relatively large, although we should keep in mind that the attenuation length will vary spatially during and after reionization, and so this quantity needs to be large only across some stretches of the IGM.
 This in turn demands some minimum separation between the neutral islands, because otherwise the neutral islands themselves will limit the mean free path and prevent a sufficiently
 intense UV radiation field from building up between the islands. Hence, it may be inconsistent to have remaining neutral islands in the IGM, yet still have some transmission through the Ly-$\alpha$ forest. Here we briefly quantify this reasoning; we will be content with only a rough estimate, as our focus
here is more on designing empirical tests. Further theoretical
exploration here might be valuable, however, perhaps along the lines of \cite{Xu:2013npa}. 

Quantitatively, previous studies infer that a photoionization rate on the order of $\Gamma_{\rm HI} \sim 5 \times 10^{-13} \text{s}^{-1}$ is required to match the mean transmitted flux in the
$z \sim 5.5$ Ly-$\alpha$ forest (e.g., \citealt{KuhlenConcordance}, \citealt{Bolton:2007b}).\footnote{Theses studies assume that reionization is complete at these redshifts. If the universe is in fact partly neutral, then a higher photoionization rate should be required in the ionized regions. In our rough estimate here, we neglect this given the other significant uncertainties involved.}
If we demand that the photoionization rate between the neutral islands needs to be in this ballpark to allow transmission through the forest, we can translate this into a required minimum average separation
between the neutral islands, given an assumed ionizing emissivity. The average ionizing emissivity is likely on the order of $\epsilon_{\rm HI} \sim 3$ photons per atom per Gyr (e.g. \citealt{Bolton:2007b}).
This is close to the value required simply to balance recombinations and maintain the ionization of the IGM at the redshifts of interest. This emissivity is also comfortable with that inferred from the above
photoionization rate and measurements of the mean free path to ionizing photons (\citealt{Bolton:2007b}, although \citealt{Becker:2013ffa} recently argued for a slightly larger value), as well as the UV emissivity implied by measurements of the galaxy luminosity function (e.g. \citealt{Robertson:2013bq}).

In this context, it is useful to note that:
\begin{align}
\Gamma_{\text{HI}} &= \varepsilon_{\text{HI}} \sigma_{\text{HI,lim}} \lambda_{\text{mfp}} \dfrac{\beta}{\beta + 1.5},
\end{align}
where $\varepsilon_{\text{HI}}$ is the average proper ionizing emissivity, $\sigma_{\text{HI,lim}} = \pow{6.3}{-18}\text{cm}^{2}$ is the photoionization cross section at the Lyman limit, $\lambda_{\text{mfp}}$ is the mean free path of ionizing photons at the Lyman limit, and $\beta$ is the intrinsic, unhardened spectral index of the ionizing radiation.  This expression assumes that the mean free path to ionizing photons
propagating through a clumpy IGM scales as $\nu^{3/2}$ \citep{Zuo}.
Inserting typical numbers we find:
\begin{align}
\Gamma_{\text{HI}} &= \pow{5.0}{-13} \text{sec}^{-1} \left[ \dfrac{\varepsilon_{\text{HI}}}{3\ \text{photons}/\text{atom}/\text{Gyr}}  \right]\\
& \times \left[ \dfrac{\beta}{2} \right] \left[ \dfrac{3.5}{1.5 + \beta} \right] \left[ \dfrac{1 + z}{6.5} \right]^{3} \left[ \dfrac{\lambda_{\text{mfp}}}{9.1 \ \text{pMpc}} \right].
\end{align}
In other words, to get transmission through the forest for plausible values of the ionizing emissivity, we require the mean separation between neutral islands to be
$\lambda_{\rm min} \gtrsim \lambda_{\text{mfp}} \gtrsim 9.1 {\text{pMpc}}$. This is a minimal requirement in that it assumes  the neutral islands set the mean free path, when in fact
Lyman limit systems and cumulative absorption in the mostly ionized gas may also play a role. On the other hand, the required minimum separation between the neutral islands would 
go down if a smaller $\Gamma_{\text{HI}}$ suffices to allow transmission through the forest, or if the ionizing emissivity is in fact higher. 
However, the mean free path to ionizing photons has
recently been measured at $z = 5.16$ to be $\lambda_{\text{mfp}} = 10.3 \pm 1.6 \text{pMpc}$ \citep{Worseck:2014fya}, only somewhat larger than our assumed $\lambda_{\text{mfp}}$ here. While there are still uncertainties, and while the measured mean free path scales steeply with redshift ($\lambda_{\text{mfp}} \propto (1+z)^{5.5}$), viable models are unlikely to have neutral islands spaced much more closely than this.

We can then use this requirement on $\lambda_{\text{mfp}}$ to get some sense of which volume-averaged neutral fractions are plausible at $z \sim 5.5$. In the simulation outputs considered here (see \S\ref{sec:Sims}), the 
mean separation between neutral islands is $\lambda_{\text{mfp}} = 17.0$ pMpc, 5.3 pMpc, and 2.7 pMpc for $\axhi = 0.05$, 0.22, and 0.35, respectively. The first case certainly satisfies
the requirement described above, the second case is just a bit on the small side, while the third case is uncomfortably small. Given the uncertainties in this argument, and the possibility that the neutral
islands are a bit larger than in our simulation (which would increase their mean separation at fixed filling factor), we consider all three cases, but refrain from considering still more neutral models. We regard the latter case ($\axhi = 0.35$) as an extreme scenario intended mostly for illustration.

Finally, it is worth keeping in mind that any remaining neutral islands will likely be photoionized on a short timescale. For example, using Eq. 1 in \citet{Lidz:2014jxa} with $C=3$, $M_{\rm min} = 10^9 M_\odot$, and
$\zeta = 20$, the redshift interval over which the volume average ionized fraction transitions from $\avg{x_i} = 0.8$ to $\avg{x_i}=1$ is only $\Delta z \sim 0.5$. 
However, it is possible that we
are catching this -- likely brief -- phase in $z \sim 5.5$ Ly-$\alpha$ forest
spectra and the possibility of testing this remains tantalizing.

\section{Simulations and Mock Spectra} \label{sec:Sims}

With the above discussion to frame the range of possibilities, we move to describe the numerical simulations used in this analysis and our approach to constructing mock Ly-$\alpha$ forest
absorption spectra before reionization completes. We use simulated density and ionization fields generated from a dark matter simulation of \cite{McQuinn:2007dy} which tracks $1024^{3}$ dark matter particles in a simulation volume with a co-moving sidelength of $L = 130\mpch$. We assume that the gas closely follows the dark matter. In this work, we focus on redshift $z = 5.5$, but consider several possible neutral fractions. In practice, we obtain ionization fields with higher (lower) neutral fractions by using simulation outputs at higher (lower) redshifts. This should be an appropriate approximation since the statistical properties of ionized regions at a given neutral fraction are most sensitive to the neutral fraction and are relatively insensitive to the redshift at which the neutral fraction was attained (see \citealt{McQuinn:2006et} and \citealt{Furlanetto:2004nh}).

We generate mock quasar spectra according to the usual ``fluctuating Gunn-Peterson" approach (e.g., \citealt{Croft:2000hs}), with a few refinements to capture the main effects of incomplete reionization. First and foremost, we do not assume a fully ionized IGM. The transmission in the Ly-$\alpha$ forest is sensitive to the precise ionized fractions in the ionized phase of the IGM. In order to simplify our study, as mentioned in the previous section, we rescale the simulated photoionization rates in the ionized regions to match the observed mean transmitted flux through the Ly-$\alpha$ forest. We do this assuming
ionization equilibrium, and a constant value of the UV background (with a photoionization rate per atom of $\Gamma_{\text{HI}}$) within ionized regions. Specifically, simulated pixels with $x_i > 0.9$ are considered highly ionized while less ionized pixels are considered fully neutral.\footnote{After effectively thresholding the ionization field in this way, we end up with neutral fractions which are $\approx 20\%$ higher than in the original simulation. Throughout the paper, we refer to increased, thresholded neutral fractions.} This simplified approach allows us to consider a range of different possibilities for
the ionization state of the IGM quickly. We comment on the shortcomings of this approach when appropriate.
The optical depth of a given pixel, $i$, in the simulation can then be found by summing over contributions from neighboring pixels (Bolton \& Haehnelt 2008):
\begin{align}
\tau_{\alpha}(i) &= \frac{c \sigma_{\alpha}\delta R}{\pi^{1/2}}\sum_j \frac{n_{\text{HI}}(j)}{b(j)}H(a,x),
\end{align}
where $b(j) = (2 k_{\text{B}}T(j)/m_{p})^{1/2}$ is the Doppler parameter, $T(j)$ is the temperature of pixel $j$, $\delta R$ is the pixel proper width, $\sigma_\alpha = \pow{4.48}{-18}\text{cm}^{2}$ is the \lya\ scattering cross section, $m_p$ is the proton mass, $H(a,x)$ is the Hjerting function, and $n_{\text{HI}}(i)$ is the number density of hydrogen atoms at pixel $i$, found using the simulated density field. To calculate the Doppler parameter, we assume that the gas obeys a modified temperature-density relationship
\begin{align}
T(\delta) = \begin{cases} T_0(1 + \delta)^{\gamma - 1} &\mbox{if ionized}\\
1,000 \text{K} &\mbox{if neutral}, \end{cases} \label{eq:TDrelation}
\end{align}
where $\delta$ is the matter overdensity in units of the cosmic mean and we choose $T_{0} = \pow{2}{4}$K and $\gamma = 1.3$ as the temperature at mean density and slope of the temperature-density relation, respectively. For simplicity, we assume the ionized gas lies on the aforementioned temperature-density relation, although there should be significant scatter around this relation close to
reionization (e.g. \citealt{Lidz:2014jxa}). We do not expect this to impact our conclusions significantly. The neutral gas should be colder than the ionized gas, of course, with a temperature set perhaps by low levels of X-ray pre-heating. Here
we adopt $T =1,000$ K for the neutral gas; this choice is likely a bit large (it was chosen partly for ease in computing the Hjerting function below), but we have checked that we get nearly identical results for colder temperature choices.

The Hjerting function is a convolution of a Lorentzian profile, which incorporates the natural line profile of the Lyman-series lines, with a Maxwell-Boltzmann distribution, which accounts for the effects of thermal broadening on the line profile. The Hjerting function is defined by:
\begin{align*}
H(a,x) &= \frac{a}{\pi} \int_{-\infty}^{\infty}\frac{e^{-y^2}\dd y}{a^{2}+ (x-y)^2},
\end{align*}
where $a = \Lambda_{\alpha} \lambda_{\alpha}/4\pi b(j)$, $\Lambda_{\alpha} = \pow{6.265}{8}\sec^{-1}$ is the damping constant, $\lambda_{\alpha} = 1215.67 \text{\AA}$ is the \lya\ wavelength, $x$ is the relative velocity of pixel $i$ and pixel $j$ in units of the Doppler parameter, defined as $x = \left[ v_{\text{H}}(i) - u(j) \right]/b(j)$, where $u(j) = v_{\text{H}}(j) + v_{\text{pec}}(j)$. The peculiar velocity field is generated by applying linear perturbation theory to the underlying density field.\footnote{This was done because the full peculiar velocity field was not readily available, but this approximation should not impact our results.}  In detail, the natural line profile is only approximately described by a Lorentzian \citep{Lee:2013fga}, with asymmetric corrections becoming important far from line center. In this study, the
precise shape of the damping wing far from line center is unimportant: we make use only of the gradual recovery in transmission around saturated neutral regions, rather than the detailed shape of this
recovery, which is also strongly influenced by neighboring neutral regions. We hence expect the Lorentzian form to be a good approximation for our present purposes.

In addition to including absorption from the hydrogen damping wing, we also include absorption from primordial deuterium. As a result of big bang nucleosynthesis, primordial hydrogen should be accompanied by traces of deuterium, with a relative abundance by number of $\text{D}/\text{H} = 2.5\times 10^{-5}$ (\citealt{Cooke:2013cba}). Due to its slightly increased reduced mass, Lyman series transitions in deuterium will be shifted blueward by 82\kms\ compared to the same transitions in hydrogen. We account for deuterium by scaling the number density of hydrogen in a given pixel by the relative abundance and shifting the resulting optical depths blueward by 82\kms. Additionally, the Doppler parameter is adjusted to $b_{\text{D}}(j) = \left( 2k_{\text{B}}T(j)/(2m_{p}) \right)^{1/2}$ to account for the increase in mass.

In this work, we focus mostly on $z=5.5$ and adopt a mean transmitted flux at this redshift of $\avg{F}=0.1$, consistent with determinations from e.g., \cite{Becker:2001ee}. In some cases, we test the
sensitivity of our results to the mean transmitted flux by considering $\avg{F}=0.05$ as well. In general, the lower the mean transmitted flux, the more challenging it is for us to identify any remaining
neutral islands. On the other hand, the likelihood that neutral islands remain increases towards high redshift and decreasing mean transmitted flux. As mentioned previously, we rescale the simulated photoionization rates in the ionized regions to a uniform value,
normalized so that an ensemble of mock spectra matches the observed mean transmitted flux. It is important to note that the mean transmitted flux is a very steep function of redshift near $z \sim 5.5$, and that the sightline-to-sightline scatter in this quantity is substantial \citep{Fan:2005es}, and so one may want to carefully test for sensitivity to the precise redshift binning used.

We use the same approach as described above to generate \lyb\ mock spectra, with $\lambda_{\beta} = 1025.72\text{\AA}$, $\Lambda_\beta = \pow{1.897}{8}\sec^{-1}$, and $\sigma_\beta = \pow{7.18}{-19}\cm^{2}$. However, in generating \lyb\ mock spectra, we must also account for foreground \lya\ absorption due to gas at lower redshifts, $\lambda_\alpha (1+z_{\text{Ly}\alpha}) = \lambda_{\beta}(1 + z_{\text{Ly}\beta})$, where $z_{\text{Ly}\alpha}$ is the redshift of the foreground \lya\ absorber and $z_{\text{Ly}\beta}$ is the redshift of the \lyb\ absorber. We will assume we are investigating quasar spectra at $z_{\text{Ly}\beta} =  5.5$ for this work, such that the corresponding foreground \lya\ absorption in the \lyb\ spectra occurs at redshift $z_{\text{Ly}\alpha} = 4.5$. We adjust $\Gamma_{\text{HI}}$ for the foreground \lya\ absorption to match measurements of the mean transmission from \cite{2013MNRAS.430.2067B} at these redshifts ($\left\langle F \right\rangle \approx 0.31$ at $z = 4.5$).\footnote{We generate foreground \lya\ absorption by considering the absorption from regions in the same simulation box, but demand that they are widely-separated from the high redshift regions of interest ($>10\mpch$). This enforces that the underlying density fields sourcing the \lyb\ absorption and the foreground \lya\ absorption are uncorrelated, as should be the case for actual spectra.} The optical depth of a pixel in a \lyb\ spectrum is then the sum of the contribution from the foreground \lya\ absorption and the intrinsic \lyb\ absorption $\tau_\beta^{\text{tot}}(z_{\text{Ly}\beta}) = \tau_{\beta}(z_{\text{Ly}\beta}) + \tau_{\alpha}(z_{\text{Ly}\alpha})$.

\begin{figure}[h]
  \centering
  \includegraphics[width=9cm]{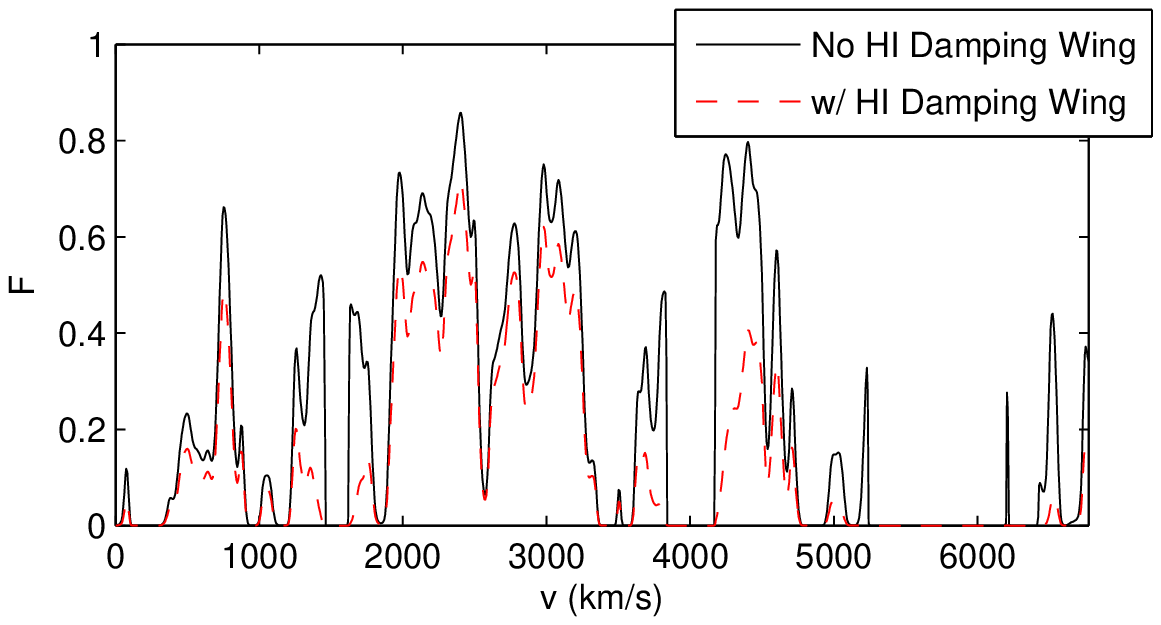}
  \includegraphics[width=9cm]{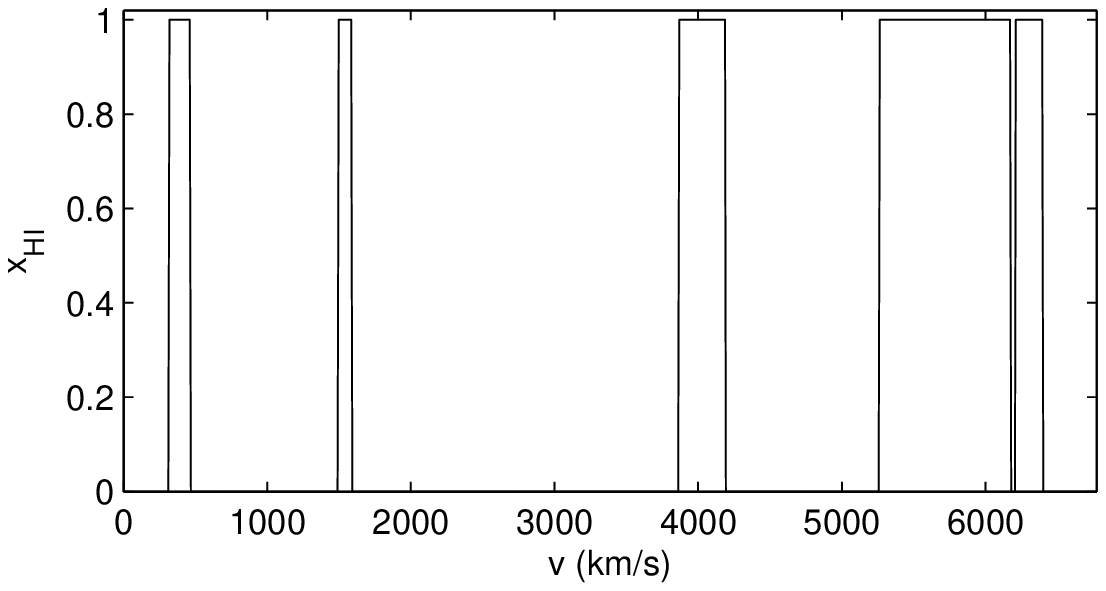}
  \caption{Example mock \lya\ forest spectrum and corresponding neutral fraction. The top panel shows the \lya\ transmission while the bottom panel is the neutral fraction along the line of sight, with ionized regions set to $x_{\rm HI} \approx 0$ for illustration. The black curve in the top panel shows the transmission through the forest when absorption due to the hydrogen damping wing is neglected, while the red curve includes damping wing  absorption. The comparison illustrates that damping wing absorption has a prominent impact, but it is also clear that the presence of the damping wing will be hard to discern by eye. The line of sight is extracted from a model with $\axhi = 0.22$, but note that we have deliberately chosen a sightline with more neutral regions than typical.}
  \label{fig:ExampleSpectraA}
\end{figure}

In \Fig{fig:ExampleSpectraA}, we show an example mock \lya\ spectrum for a particular line of sight through the simulation. We show only a portion of the line of sight in order to exhibit smaller-scale features. The top figure shows the \lya\ transmission when the hydrogen damping wing is neglected (black) and when it is included (dashed red), while the bottom panel shows the underlying thresholded ionization field. We have neglected peculiar velocities in creating this figure in order to facilitate a comparison between the spectrum and the underlying ionization field.

From this figure, we see that the damping wing indeed has a significant effect on the transmission, but that its effect is hard to discern without knowing the damping-wing-less transmission. This is the case for two reasons. First, the forest here is very absorbed and the damping wing absorption becomes mixed with resonant absorption from neighboring ionized regions. Second, the damping wing from a particular neutral region may overlap with the damping wing from another neutral region, altering the shape of the resulting absorption. Specifically, we see that, in the example spectra, the region at $v \approx 4500\ \kms\ $ is sandwiched between HI regions to the left and right, both within 1000\kms. Therefore, the optical depths for the corresponding pixels likely have significant contributions from resonant absorption, damping wing absorption from the HI region to the left, and damping wing absorption from the HI region to the right. While detecting individual instances of damping wing absorption in this case 
seems impossible, we will show that detecting the presence of damping wing absorption \textit{on average} should be feasible through the stacking of high-redshift quasar spectra.

\section{Dark Gap Statistics} \label{sec:HIDistributions}

With the mock spectra of the previous section in hand, we now consider the size distribution of regions of saturated absorption -- dark gaps -- and its dependence on the underlying neutral fraction. Using such dark gap statistics has been widely discussed as a potential probe of the IGM neutral fraction (see, e.g., \citealt{Fan:2005es}, \citealt{2010MNRAS.407.1328M}, \citealt{McGreer:2011dm}). In a fully ionized IGM, the size of dark gaps in quasar spectra should grow with increasing redshift, simply owing to the increasing mean density of the universe and as a result of any decline in the intensity of the UV radiation background. However, once quasar spectra start to probe the tail end of reionization, the increase in dark gap size should accelerate due to the presence of islands of neutral hydrogen. 

\begin{figure}[h]
  \centering
  \includegraphics[width=9cm]{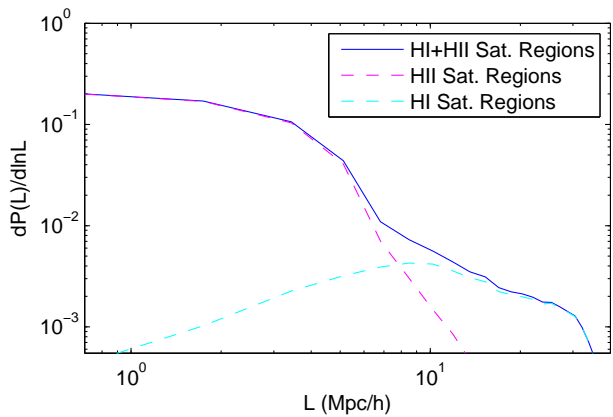}
  \caption{Dark gap size distribution for the $\axhi = 0.22$, $\langle F \rangle = 0.1$ model. The solid blue curve shows the total distribution of dark gaps from an ensemble of mock spectra, where the magenta (cyan) curve shows the same thing but for the dark gaps sourced by ionized (neutral) gas. Here, we have focused on dark gaps with $L > 0.75 \mpch$. This clearly demonstrates that neutral hydrogen is the dominant source of {\em large} dark gaps in our mock spectra, provided there is an appreciable neutral fraction.}
  \label{fig:DarkGapHist}
\end{figure}

In \Fig{fig:DarkGapHist}, we have plotted the size distribution of dark gaps, $\text{d}P(L)/\text{d}\ln L$, in blue for the $\axhi = 0.22$ mock spectra, assuming a mean transmission of $\langle F \rangle = 0.1$. Additionally, the dashed curves display the two underlying populations of dark gaps: those sourced by ionized gas (magenta) and those sourced by neutral gas (cyan). For clarity, we only show dark gaps larger than $L_{\text{sat,min}} = 0.7\mpch$ ($\sim 90 \kms$), since smaller saturated regions will be predominantly ionized. Additionally, we have neglected peculiar velocities when generating spectra here. Two important points become apparent from this figure. First, at $ L \sim 8.5\mpch$ ($\sim 1100 \kms$), dark gaps transition from being primarily sourced by ionized gas to being primarily sourced by neutral gas. This reinforces our intuition that, in a partially neutral IGM, the largest dark gaps should correspond to the remaining neutral islands. Second, the dark gaps being composed of two different populations gives rise to a bimodality in the size distribution. This suggests that the behavior of the large-$L$ tail of the size distribution may offer additional information about the neutral fraction, with a steep decline suggesting a highly ionized IGM and a more gradual decline, or the emergence of a second peak, suggesting a significantly neutral IGM. Such a ``knee" in the dark gap size distribution is also mentioned in \cite{2010MNRAS.407.1328M}.

\begin{figure}[h]
  \centering
  \includegraphics[width=9cm]{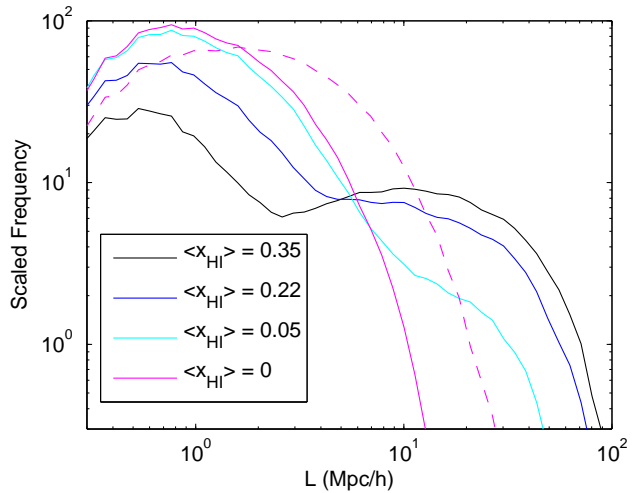}
  \caption{Large-length tail of the dark gap size histogram for $\axhi = 0$ (magenta), 0.05 (cyan), 0.22 (blue), and 0.35 (black) for the case when $\left\langle F \right\rangle = 0.1$. The y-axis is scaled to indicate the expected number of dark gaps obtainable from 20 spectra. Bins in this figure are spaced logarithmically. The dashed magenta line indicates the dark-gap size distribution in the fully ionized case when the true transmission is $\left\langle F \right\rangle = 0.03$, but continuum fitting errors result in a measured mean transmission of $\left\langle F_{\text{meas}} \right\rangle = 0.1$.}
  \label{fig:PDFs}
\end{figure}

Additionally, we can consider the large-$L$ tail of the size distribution and its dependence on neutral fraction at a fixed mean transmission. In \Fig{fig:PDFs}, we plot an expected histogram of dark gap sizes for 10 spectra for  $\axhi = 0$ (magenta), 0.05 (cyan), 0.22 (blue), and 0.35 (black), again assuming that $\langle F \rangle = 0.1$. Three trends become obvious from this plot. First, as the neutral fraction is increased (at fixed $\langle F \rangle $), the number of large saturated regions increases and, second, as the neutral fraction is increased, the size of the largest dark gaps also increases. For example, in the $\axhi = 0.22$ model, the largest dark gaps are roughly five times bigger than in the fully ionized model. Additionally, we again see hints of the underlying dark gap size distribution being bimodal as the neutral fraction is increased, supporting the idea that the \textit{shape} of the dark gap size distribution may be a diagnostic for the underlying neutral fraction.

Given these trends, it should be possible to compare dark gap distributions from observed spectra against models at various neutral fractions and use this to constrain the mean neutral fraction of the IGM. This approach is appealing in that it does not require especially high-resolution or high signal-to-noise spectra. However, it does require comparison with simulated models 
of the dark gap
size distribution and so the conclusions reached will be somewhat model dependent.
Additionally, the distributions are dependent on the assumed mean transmission, which is itself uncertain. In particular, estimates of the mean transmission at high redshift may be impacted by continuum
fitting errors, given the inherent difficulty in estimating the unabsorbed continuum level in highly-absorbed spectra.

In order to investigate the impact of possible continuum fitting errors, we generate mock spectra in the fully ionized model with $\avg{F}=0.03$ but then rescale the flux in each simulated
pixel by a multiplicative factor -- to mimic the effect of continuum misplacement -- such that the measured mean transmitted flux appears to be $\avg{F_{\text{meas}}} = 0.1$. This case is shown as
the magenta dashed line in \Fig{fig:PDFs}. Here the dark gap size distribution is shifted towards sizes than one would expect in an ionized model at $\avg{F}=\avg{F_{\text{meas}}} = 0.1$.
However, the shape of the size distribution is still quite different than in the partly neutral models. Importantly, the dark gap distribution in the ionized model still lacks the distinctive bump at large
sizes that is the hallmark of a partly neutral IGM in our models.

\section{Stacking Toy Spectra} \label{sec:Stacking}

In this section we describe our basic approach of stacking \lya\ and \lyb\ spectra in order to detect the presence of HI damping wings and absorption due to deuterium, respectively. While the forest is too absorbed at these redshifts to easily detect damping wings or deuterium absorption due to individual neutral regions, here we demonstrate that the presence of such features can be revealed \textit{on average} by stacking regions of transmission over many spectra.

This section serves as a proof of principle by applying a simplified stacking approach to mock spectra generated using an idealized IGM model. Specifically, we consider an ensemble of sightlines through our simulation box and assume that the IGM is entirely ionized with the exception of a single HI island with mean density and varying length, $L$, inserted randomly along each line of sight. We then generate mock spectra assuming these density and ionization fields. The stacking in this section is always done starting at the HI/HII boundaries of a given HI region moving outward.

\subsection{HI Damping Wing} \label{sec:ToyHI}

Our stacking approach can be clearly demonstrated by considering the damping wing from neutral hydrogen. Due to the natural width of the \lya\ line, a neutral hydrogen gas parcel should cause \lya\ absorption over a range of frequencies. Far from line center, this absorption will have an optical depth roughly following (\citealt{MiraldaEscude:1997en}):
\begin{align}
\tau_{\text{Ly}\alpha}^{\text{DW}}(\Delta v) \approx \frac{\tau_{\text{GP}}R_{\alpha}c}{\pi}\left[ \frac{1}{\Delta v} - \frac{1}{\Delta v + v_{\text{ext}}} \right]  \label{eq:HIDW}
\end{align}
where $\tau_{\text{GP}}$ is the Gunn-Paterson optical depth, $R_{\alpha} \equiv \Gamma_{\alpha}\lambda_{\alpha}/4\pi c$, $\Gamma_{\alpha} = \pow{6.265}{8}\sec^{-1}$ is the \lya\ decay constant, $\Delta v$ is the separation from the HI/HII boundary in velocity space, $v_{\text{ext}}$ is the extent of the hydrogen region in velocity space, and $c$ is the speed of light. For a large neutral region, this equation implies that $\tau^{\text{DW}}_{\text{Ly}\alpha}(|v| < 600\kms) \geq 1$ at $z \sim 5.5$. This excess absorption is referred to as the hydrogen ``damping wing". While both neutral gas and highly ionized gas can cause absorption in quasar spectra, only a significantly neutral hydrogen patch will result in damping wing absorption, owing to the greatly reduced optical depth in the wing compared to line center. As such, detecting damping wing absorption would be a smoking gun for the presence of significantly neutral hydrogen islands. Note that the transmission profile will differ from the simple form of Eq. \ref{eq:HIDW},
owing mostly to neighboring neutral regions, however the gradual recovery to transmission around saturated neutral regions should be a distinctive indicator that highly neutral regions remain in the IGM.

\begin{figure}[h]
  \centering
  \includegraphics[width=8.4cm]{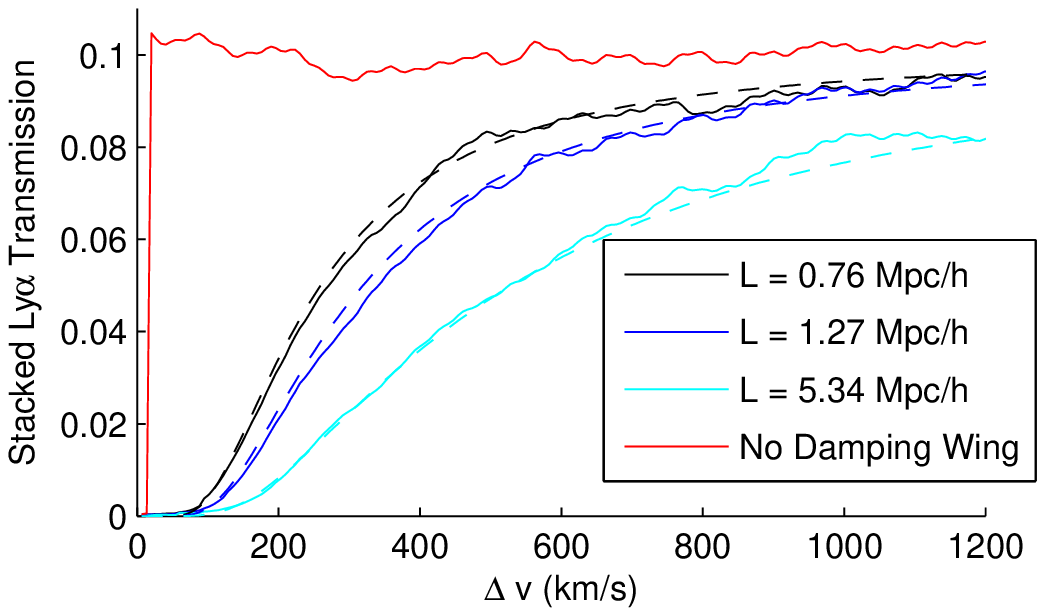}
  \caption{Stacking idealized \lya\ spectra containing toy HI regions. The above figure shows the stacked transmission outside isolated HI regions with mean density and size $L = 0.76 \mpch$ ($\vext \approx 100\kms$), $L = 1.27 \mpch$ ($\vext \approx 170\kms$), and $L = 5.34 \mpch$ ($\vext \approx 700\kms$) shown in black, blue, and cyan, respectively. The solid red curve shows the stacked transmission outside of the same HI regions \textit{neglecting} the damping wing, which will be the same on average in all cases. In generating these spectra, we assume $\left\langle F \right\rangle = 0.1$. In this greatly-idealized case, the presence of the hydrogen damping wing is seen clearly through extended excess absorption compared to the red curve. Furthermore, we can see that the excess absorption closely follows what we would expect analytically based on multiplying \Eqn{eq:HIDW} by the overall mean transmission. In this figure, all stacking starts at HI/HII boundaries.}
  \label{fig:ToyHI}
\end{figure}

In \Fig{fig:ToyHI} we show the results of stacking transmission outside of neutral regions in the toy mock spectra described earlier in this section, neglecting deuterium for the time being. Namely, we show the stacked transmission outside neutral islands of length $L = 0.76 \mpch$ ($\sim 100\kms$) in black, $L = 1.27 \mpch$ ($\sim 170\kms$) in blue, $L = 5.34\mpch$ ($\sim 700\kms$) in cyan, and stacked transmission neglecting the damping wing in red. Additionally, we have plotted the analytic curves corresponding to \Eqn{eq:HIDW} for the various $L$ values, shown with dashed curves. We have applied a single multiplicative factor to these curves to account for average resonant absorption from ionized gas. Together, this figure implies that damping wing absorption from isolated neutral regions has a significant impact on quasar spectra, extending $\sim 1000\kms\ $ past the HI/HII boundaries, which may be observable through stacking as expected from Eq. \ref{eq:HIDW}. 

In providing a toy example of how the hydrogen damping wing can affect spectra, we have neglected many important challenges that such a measurement would face. For example, we assumed perfect knowledge of the underlying ionization state of the IGM in order to determine where to stack and we assumed that we could discriminate between neutral and highly ionized absorption systems. However, the presence of such a large and potentially-observable feature provides motivation for us to apply the stacking approach in a more realistic manner. In \S \ref{sec:RealSpectra} and in Appendix B, we describe several such challenges and subtleties along with potential resolutions.

\subsection{Deuterium} \label{sec:ToyD}

With the stacking approach of the previous section in mind, we now consider absorption due to deuterium. As noted in \S\ref{sec:Sims}, primordial hydrogen should be accompanied by traces of deuterium, with a relative abundance of $\sim 2.5\times 10^{-5}$ (\citealt{Cooke:2013cba}). Due to its slightly increased reduced mass, atomic transitions in deuterium are shifted blueward by 82\kms\ compared to the same transitions in hydrogen. This implies that absorption due to neutral hydrogen in the IGM should be accompanied by additional absorption from deuterium, shifted blueward by 82\kms. We can estimate the optical depth for \lya\ absorption in deuterium at cosmic mean density by simply scaling the hydrogen \lya\ optical depth by the deuterium abundance:
\begin{align}
\tau_{D,\alpha} &= \left[ \frac{D}{H} \right]  \times \tau_{\text{GP}} \approx 8.25 x_{\text{HI}} \left(1+\delta\right)  \left[ \frac{1+z}{6.5} \right]^{3/2}.
\end{align}
Thus, we see that while the relative abundance of deuterium is extremely small, the Gunn-Peterson optical depth is so large that the resulting deuterium optical depth is still of order 10 in \lya.

An appealing aspect of searching for damping wing absorption is that the optical depth in the wing is large enough to cause significant absorption in the presence of neutral islands, but small enough to be negligible for ionized absorption systems. We see this again in the case of deuterium absorption, suggesting that it may be useful as an additional ``smoking gun" indicator for underlying neutral hydrogen. However, an obvious problem with detecting deuterium in \lya\ spectra is that the feature should be narrow and well within the broad range of velocities where the hydrogen damping wing is significant. Specifically, according to \Eqn{eq:HIDW}, at $\Delta v = 82\kms$, the damping wing optical depth for an extended neutral region should be $\tau_{\text{Ly}\alpha}^{\text{DW}}(\Delta v \approx 82\kms) \approx 8$. Therefore, the feature should be completely wiped out in \lya\ spectra by the hydrogen damping wing.

However, the damping wing optical depth in the \lyb\ line is much smaller. Specifically, according to \Eqn{eq:HIDW}, the damping wing optical depth scales as 
\begin{align}
\frac{\taudwb}{\taudwa} &= \frac{\tau_{\text{GP},\beta}}{\tau_{\text{GP},\alpha}} \times \frac{R_{\beta}}{R_{\alpha}} = \frac{f_{\beta}\lambda_{\beta}}{f_{\alpha}\lambda_{\alpha}} \times \frac{(\Gamma_{\beta}+\Gamma_{\text{H}\alpha})\lambda_{\beta}}{\Gamma_{\alpha}\lambda_{\alpha}} \nonumber \\
&= \dfrac{f_{\beta}^{2}}{f_{\alpha}^{2}} \left( 1 + \dfrac{f_{\text{H}\alpha}}{f_{\beta}} \dfrac{\lambda_{\beta}^2}{\lambda_{\text{H}\alpha}^2} \right) = .0410,
\end{align}
and should therefore be significantly narrower in \lyb\ than in \lya. In the above expression, $f_{\alpha}$, $f_{\beta}$, and $f_{\text{H}a}$ are the oscillator strengths of the \lya, \lyb, and Balmer-$\alpha$ transitions, respectively, with $\lambda$ denoting the corresponding wavelengths and $\Gamma$ denoting the corresponding decay constants. By modifying \Eqn{eq:HIDW} for \lyb, we see that $\taudwb(|\Delta v| \gtrsim 25\kms) \leq 1$. Therefore we find that \textit{the hydrogen damping wing should not wipe out deuterium absorption features in } \lyb. Furthermore, while the hydrogen damping wing optical depth is reduced by a factor of roughly $f_{\beta}^{2}/f_{\alpha}^{2}$ when considering \lyb, the total optical depth in the deuterium line is only reduced relative to deuterium \lya\ by $f_{\beta}\lambda_{\beta}/f_{\alpha}\lambda_{\alpha} \approx 1/6$, such that the optical depth should still be of order 1 for deuterium \lyb. Therefore, not only should a deuterium absorption feature survive the hydrogen damping wing, but it should still have a strong enough optical depth to cause significant absorption if neutral islands in fact remain.

Naturally, it should be very difficult to detect individual deuterium absorption features from the diffuse IGM, as the Ly-$\beta$ spectra will be very absorbed when the universe is neutral enough to
produce the features in the first place. However, the feature may nonetheless be observable \textit{on average} through the stacking of high-resolution quasar spectra. 
In order to demonstrate the strength of the deuterium absorption feature in stacked spectra, we incorporate deuterium into the same toy sightlines from \S \ref{sec:ToyHI} to produce mock \lyb\ spectra, neglecting foreground \lya\ absorption for the time being. We are then able to stack transmission outside of neutral regions in the spectra, starting at the HI/HII boundaries and moving outward. However, since deuterium absorption will only occur on the blue side of neutral regions, we need only stack those regions of transmission. In fact, this offers a clean test for detecting deuterium. Namely, we can separately stack transmission redward and blueward of neutral regions and compare. Excess absorption on the blue side of neutral regions, on average, could signal the presence of deuterium absorption. This is especially appealing since there should be no sources of contamination that would cause a similar, and significant, red/blue asymmetry.\footnote{One source of asymmetry we do find, which can be seen in \Fig{fig:LybResults}, results from the fact that, when dealing with realistic spectra, we force there to be transmission in \lyb\ at the locations where stacking begins. This results in a small selection effect, where selected neutral absorption systems have a reduced probability of having nearby neutral regions and have correspondingly-smaller nearby optical depths. For deuterium, this smaller optical depth is shifted blueward, causing \textit{less} absorption on the blue side of the line for $\Delta v \gtrsim 82\kms$. However, this asymmetry is minor and {\it opposes} the asymmetry from deuterium absorption.}

\begin{figure}[h]
  \centering
  \includegraphics[width=8.4cm]{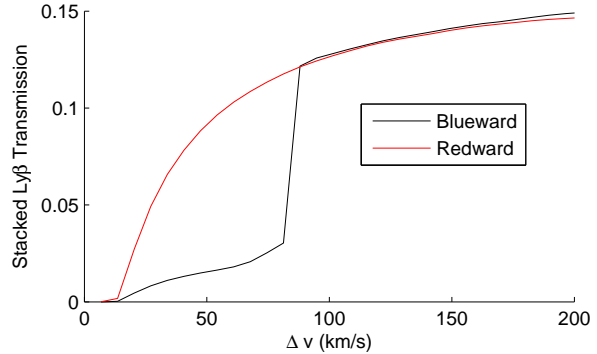}
  \caption{Presence of deuterium absorption revealed through stacking idealized \lyb\ spectra containing toy neutral regions. The red and black curves show the stacked \lyb\ transmission redward and blueward, respectively, of toy neutral regions of length $L = 5\mpch$ ($\approx 700\kms$) randomly inserted into many sightlines, with spectra generated assuming $\left\langle F_{\text{Ly}\alpha} \right\rangle = 0.1$. In each case, stacking begins at the underlying HI/HII boundary. We have also mimicked the effect of including foreground \lya\ absorption by scaling the feature by the mean transmission in the foreground \lya\ forest. This demonstrates that, at least in this idealized case, the presence of deuterium absorption can be easily seen out to $\sim80\kms$\ past the HI/HII boundaries.}
  \label{fig:ToyD}
\end{figure}

In \Fig{fig:ToyD}, we show the results of stacking transmission in \lyb\ redward (red) and blueward (black) of the toy neutral regions across the full ensemble of mock quasar spectra. As in \Fig{fig:ToyHI}, all stacking begins at HI/HII boundaries. We can see the blueward transmission clearly exhibits excess absorption due to deuterium extending roughly $\sim$80\kms\ from the HI/HII boundary. Thus, in this idealized scenario, the presence of deuterium in islands of neutral hydrogen leaves a very clear signature in the stacked \lyb\ transmission.

Before proceeding further, we should point out one important caveat here. In our simulated models, the transition between fully neutral and highly ionized regions is, by construction, perfectly sharp. If this
transition is more gradual in reality, then the narrow deuterium feature could be overwhelmed by absorption from mostly ionized hydrogen in this transition region. A minimal scale for this transition
region is set roughly by the mean free path to ionizing photons through the neutral IGM, which is only $\lambda_{\text{HI}} \sim 1/(n_{\text{HI}} \sigma_{\text{HI}}) \approx 6 \text{proper kpc}/h \approx 0.8 \kms$.
This minimal scale is two orders of magnitude smaller than the scale of the deuterium feature and hence does not present a worry. However, if the edges of the ionized regions tend to experience
a reduced ionizing background, this might obscure the deuterium feature, even in the case of a partly neutral IGM. We believe the possibility of detecting this deuterium feature is enticing enough to
warrant further investigation.

As was the case in \S \ref{sec:ToyHI}, we have made several simplifying assumptions and have additionally neglected foreground \lya\ absorption from the lower-redshift IGM. However, the clear presence of deuterium absorption revealed through the simplified stacking approach provides motivation to also consider applying it to more realistic spectra, as will be discussed in \S \ref{sec:RealSpectra}.

\section{Steps of Approach} \label{sec:RealSpectra}

In \S\ref{sec:Stacking}, we demonstrated the utility of stacking idealized quasar spectra in order to reveal the presence of the HI damping wing and deuterium absorption. The success of this approach in the toy case provides motivation for us to apply it to realistic mock spectra. In doing so, we must confront the simplifying assumptions made in \S\ref{sec:Stacking}.

The most obviously unrealistic assumption made in \S\ref{sec:Stacking} is that we can precisely identify the HI/HII boundaries underlying our spectra. In practice, we will only have access to the level of transmission at each point along the spectra. However, based on \Fig{fig:ToyD}, the recovery from saturated absorption to transmission occurs within $\lesssim 15\kms$ in \lyb\ from the edge of the neutral zone, and should therefore provide a relatively
good indicator of the HI/HII boundary. Therefore, we choose to identify stacking locations based on where transmission recovers \textit{in \lyb}.
To be clear, for the case of the hydrogen damping wing, we are stacking transmission in the \lya\ forest, but we are choosing where to start the stacking based on features in the \lyb\ forest. A drawback of this approach, when searching for the hydrogen damping wing, is that we are only able to stack regions of the \lya\ forest with corresponding regions in the \lyb\ forest that are not contaminated by \lyc\ absorption. This effectively reduces the amount of usable spectra, since, for a quasar at $z = 5.5$, the pure \lya\ forest will extend $4.5 \leq z \leq 5.5$, but \lyc\ absorption will contaminate the \lyb\ forest at $z \lesssim 5.16$. If presented with a limited number of spectra, it may be worth searching for the damping wing by using only the \lya\ regions of the spectra.

By stacking at the precise locations of HI/HII boundaries in \S\ref{sec:Stacking}, we were also ensuring that our sample of absorption systems was all neutral. However, when we modify our approach to begin stacking at locations where transmission recovers from saturated absorption, we may start stacking transmission outside of ionized absorption systems together with transmission outside of neutral absorption systems, diluting our signal. Since the signal we are aiming to find is small to begin with, it is important that we minimize this contamination from ionized regions. To do this, we take advantage of the main argument of \S\ref{sec:HIDistributions}, namely that regions of saturated absorption sourced by neutral gas should be significantly larger, on average, than those sourced by ionized gas. Therefore, we choose to stack only transmission outside of \textit{large} regions of saturated absorption. Furthermore, since true neutral regions should cause saturated absorption in \lyb, we choose to stack only outside of large saturated regions which are fully absorbed in \lyb, where we define ``large" to be $>500\kms\ $ ($\gtrsim 4\mpch$) in \lyb. Note that this choice is tuned for
the case of $\avg{F} = 0.1$: a different choice may be better for other values of the mean transmitted flux. At any rate, in applying these tests to real data, one would likely vary this size scale across a range
of possible values.

Additionally, an appealing feature of the search for deuterium absorption is that it offers a very clean test for its detection, namely a red/blue asymmetry in the transmission outside of plausibly neutral regions. The disparity in the size distribution of saturated regions sourced by neutral and ionized gas suggests a similar test may be possible for the detection of the HI damping wing. Namely, while large regions of saturated absorption are likely to be sourced by neutral gas, small regions of saturated absorption are likely to be sourced by ionized gas. Therefore, to find evidence of excess absorption outside of neutral regions due to the HI damping wing, we compare the stacked transmission outside of large absorption systems, plausibly sourced by neutral gas, to that outside of small absorption systems, likely sourced by ionized gas. A significant amount of excess absorption outside of the former compared to the latter, extending further than any possible density correlations, would suggest the presence of damping wing absorption.

Furthermore, in \S\ref{sec:ToyD}, we discussed how the damping wing is greatly weakened in \lyb\ compared to in \lya. Therefore, an additional test for the presence of damping wing absorption could be to take the ratio of the stacked \lyb\ transmission to the stacked \lya\ transmission, where stacking occurs in the same physical regions in both cases. In the event that there is significant damping wing absorption, this ratio should also slowly recover to some constant value at large $\Delta v$. We further discuss and develop this approach in Appendix B.

When dealing with realistic spectra, we must adjust our approach to accommodate the presence of noise (and finite spectral resolution). While noise should average out in stacked regions, the presence of noise will also obfuscate the precise boundaries between saturated absorption and transmission. We choose to handle this by smoothing our noisy spectra over a scale of 100 km/s ($\sim 0.75\mpch$) and defining any pixel, $i$, with transmission $F_i < 3\tilde{\sigma}_{\text{N}}$ to be consistent with saturated absorption, where $\tilde{\sigma}_{\text{N}}$ denotes the standard deviation of the smoothed noise. We then define regions in the smoothed spectra where the flux goes from $F<3\tilde{\sigma}_{\text{N}}$ to $F>3\tilde{\sigma}_{\text{N}}$ as the transitions from saturated absorption to transmission, and therefore as potential points to start stacking. When stacking transmission, however, we stack the transmission in the \textit{unsmoothed} spectra.

Another concern is that damping wing absorption sourced by DLAs may erroneously be attributed to a significantly neutral IGM. However, in Appendix A, we estimate the expected rate of DLAs occurring in $z \sim 5.5$ quasar spectra and find it is small enough to be ignored. Additionally, DLAs may be discriminated from diffuse neutral islands based on the presence of metal lines and the relative sizes of their absorption in \lya\ and \lyb. 

Finally, as mentioned previously, we approximate the ionizing background in the ionized regions as uniform and ignore scatter in the temperature density relation. Accounting for these fluctuations {\em might}
lead to a more gradual recovery in the transmission around absorbed regions -- in the case of a fully ionized universe -- than in our models. Further investigation of this issue would certainly be required if a gradual recovery
is indeed found in real spectra. In Appendix B, we discuss a possible empirical test that may help in this regard.

\section{Results} \label{sec:Results}

Having considered the subtleties of the previous section, we are now ready to apply the three-pronged approach to more realistic mock spectra. In each section, we first consider the ideal case where no noise has been applied to give an idea of the potential constraining power of the different methods. Subsequently, we add realistic levels of noise and consider realistic spectra resolution to give an idea of the constraining power of the approaches applied to Keck HIRES spectra for the deuterium feature and damping wing, and spectra with slightly higher resolution than SDSS for the dark gap size distribution.

\subsection{Detecting the Damping Wing} \label{sec:Damping Wing Results}

We first consider the ability to uncover the presence of the hydrogen damping wing by strategically stacking regions of transmission in the the \lya\ forest of $z \approx 5.5$ noiseless mock quasar spectra. As discussed in \S \ref{sec:RealSpectra}, our aim is to compare the average transmission outside of plausibly neutral absorption systems to the transmission outside of likely ionized systems.

We identify the plausibly neutral absorption systems by requiring the regions be completely absorbed in \lyb, and also that the regions of saturated absorption are at least $L_{\text{sat}} > 500\kms$ ($\sim4\mpch$) \textit{in \lyb}. We begin stacking at the point in the \lya\ spectrum which corresponds to the recovery from absorption to transmission \textit{in \lyb}. We identify the likely ionized absorption systems by requiring that they are \textit{below} a maximum length $L_{\text{max}} = 300\kms$ \textit{in \lyb}.

\begin{figure}[h]
  \centering
  \includegraphics[width=9cm]{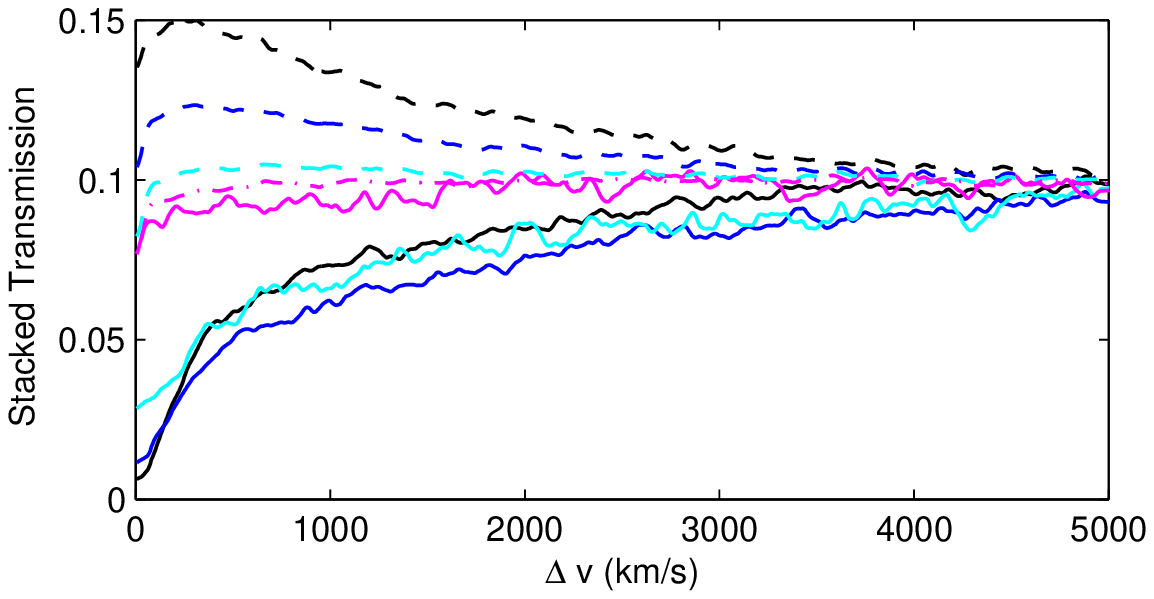}
  \includegraphics[width=9cm]{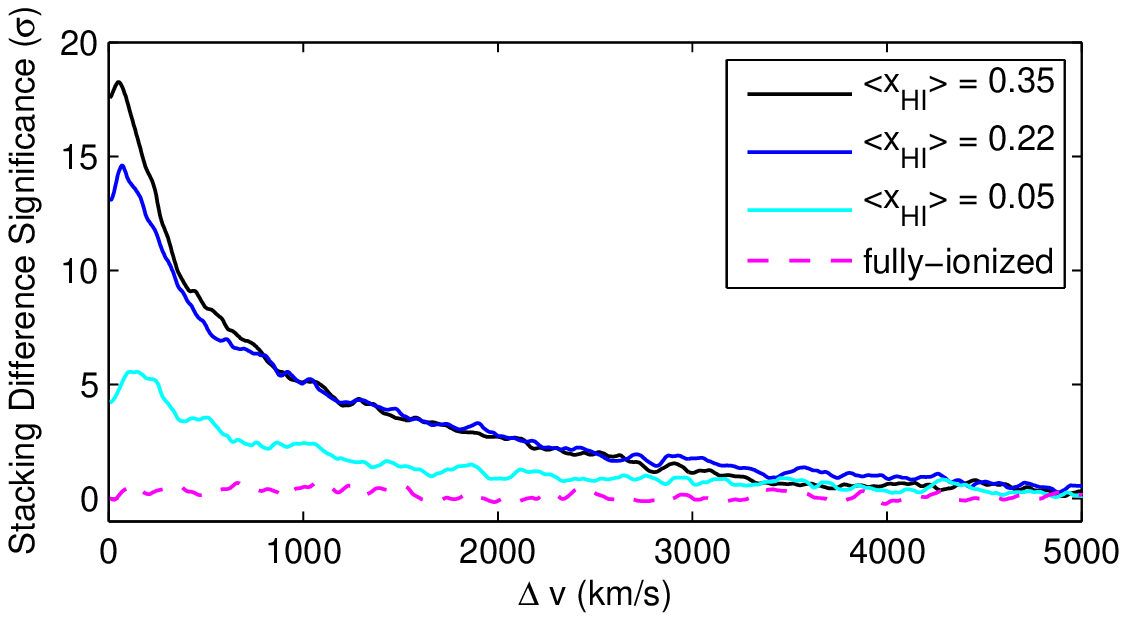}
  \caption{\lya\ stacking results for various neutral fractions. The top panel shows the mean (noiseless) stacked transmission outside of large absorption systems (solid) and small absorption systems (dashed) in the \lya\ forest for neutral fractions $\axhi = 0.35$ (black), 0.22 (blue), 0.05 (red), and 0 (magenta). The transmission here is estimated from a large ensemble of mock spectra to obtain a smooth estimate
  of the average transmission around saturated regions in each model. 
  The bottom panel shows the statistical significance of the difference between the dashed and solid curves in the top panel assuming a sample of 20 spectra are used in the stacking process.}
  \label{fig:LyaResults}
\end{figure}

In \Fig{fig:LyaResults}, we show the results of applying this approach to realistic mock spectra generated assuming various ionization states of the IGM. In the top panel, we show the stacked transmission outside of plausibly neutral absorption systems (solid) and likely ionized absorption systems (dashed), using a volume-averaged neutral fraction of $\axhi = 0.35$ (black), 0.22 (blue), 0.05 (cyan), and $\axhi = 0$ (magenta). The curves agree with our expectations, namely that transmission outside of neutral regions should recover more slowly and exhibit a rough damping wing shape with a large extent in velocity space. We see that the excess absorption extends \textit{farther} than the $\lesssim 1000\kms\ $expected from an isolated damping wing. However, as discussed in Appendix C, we find that the spatial clustering of neutral regions is responsible for this effect.

While, for several reasons discussed earlier, the shape of the absorption is distorted compared to \Fig{fig:ToyHI}, it can be seen for all significantly neutral ionization states. An important check is to apply the stacking procedure to a fully ionized IGM and ensure that we do not make a false detection. The results of this check are shown by the magenta curves in \Fig{fig:LyaResults}. As we can see, the resulting stacked transmission outside of plausibly neutral regions lacks an overall damping wing shape and stays roughly fixed near the mean transmission.

We can also see that the transmission outside of small absorption systems is very sensitive to the underlying neutral fraction. We expect this, however, since this stacked transmission depends strongly on the average transmission in regions which are not in saturated absorption, denoted $\langle F | F > 0 \rangle$. Since the dark pixel covering fraction in our mock spectra increases with $\axhi$, mock spectra with larger neutral fractions must have larger values for $\langle F | F > 0 \rangle$ to maintain $\langle F \rangle = 0.1$. As such, \Fig{fig:LyaResults} shows that the stacked transmission outside of small absorption systems increases monotonically with $\axhi$.

We estimated the stacked transmission from a large ensemble of simulated spectra to produce a smooth estimate of the average transmission around saturated regions in each model.
The transmission curves outside of individual absorption systems are, however, quite noisy on their own such that, from saturated region to saturated region, there is significant scatter about the mean-value curves shown in the top panel. In order to estimate how confidently we can distinguish the solid and dashed curves with a reasonable number of quasar spectra, we scale the number of identified absorption systems to what we would expect using $\sim 20$ spectra. Specifically, we take the difference between the dashed and solid curves and divide by the scatter in each bin. The scatter of each bin is simply the scatter in stacked transmission outside of large absorption systems, scaled by $1/\sqrt{N_{\text{sat,large}}}$, added in quadrature with the scatter in the stacked transmission outside of small absorption systems, scaled by $1/\sqrt{N_{\text{sat,small}}}$.  Here we scale to estimate the plausible scatter around the mean after estimating the transmission around saturated regions using $20$ quasar absorption
spectra.

The results of this are shown in the bottom panel of \Fig{fig:LyaResults} for the same ionization states. The results appear to be very encouraging, indicating that, assuming noiseless spectra, the solid and dashed curves are $\gtrsim 5\sigma$ statistically-significantly different (\textit{even for $\axhi = 0.05$!}). In addition, we see that the difference roughly follows a damping wing shape and remains significant for $\gtrsim 3000\kms$. We should emphasize that, while the deuterium absorption feature will necessarily be a $\lesssim 80\kms\ $ feature and require high resolution spectra to be seen, the damping wing feature extends an order of magnitude farther in velocity space and should be accessible to lower-resolution spectra.

\begin{figure}[h]
  \centering
  \includegraphics[width=9cm]{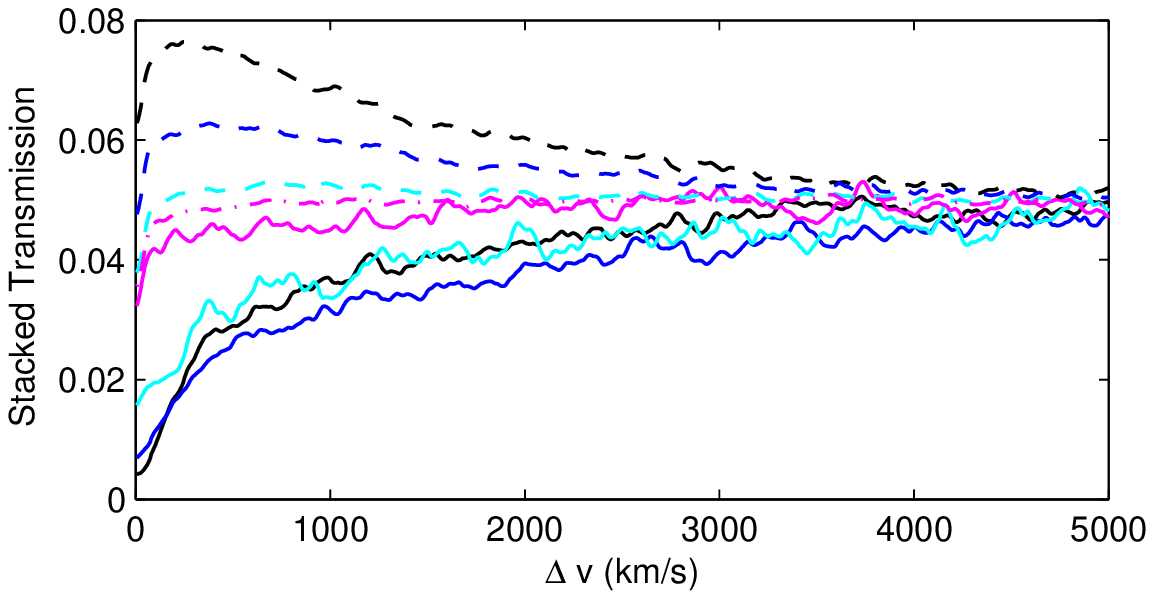}
  \includegraphics[width=9cm]{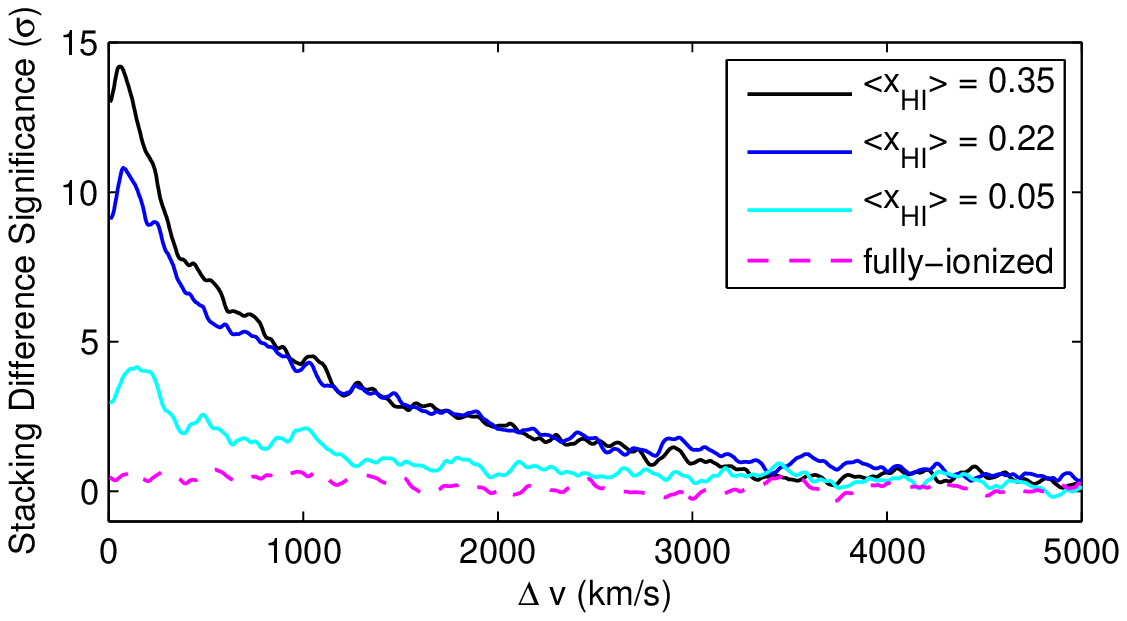}
  \caption{\lya\ stacking results assuming $\left\langle F \right\rangle = 0.05$. The above panels are identical to those in \Fig{fig:LyaResults} except that mock spectra have been generated assuming $\left\langle F \right\rangle = 0.05$. }
  \label{fig:LyaResults_LowF}
\end{figure}

In \Fig{fig:LyaResults_LowF}, we show the same results as in \Fig{fig:LyaResults}, but assume a lower mean transmission of $\left\langle F \right\rangle = 0.05$, consistent with spectra at $5.7 \lesssim z < 6$ (\citealt{Becker:2001ee}). From the figure, we see that these results are very similar to those for $\left\langle F \right\rangle = 0.10$, but with the significance curves peaking at a $\sim 70\%$ lower value and with the stacked transmission recovering to a lower mean. Overall, this provides encouragement for applying the approach to higher-$z$ spectra, suggesting that a range of physically interesting neutral fractions could be probed.

\begin{figure}[h]
  \centering
  \includegraphics[width=9cm]{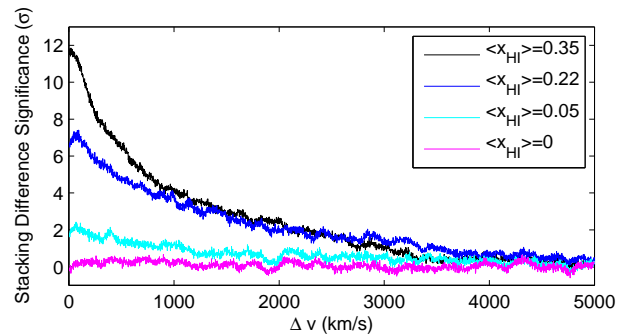}
  \caption{Results of \lya\ stacking with HIRES-style spectra ($\left\langle F \right\rangle = 0.1$). The above panel is identical to the bottom panel in \Fig{fig:LyaResults} except that the spectra have had the bin size and spectral resolution adjusted to match that of Keck-HIRES spectra. Additionally, we have added noise such that the spectra have a signal-to-noise value of 10 per pixel at the continuum.}
  \label{fig:HIRES_LyaResults_Noisy}
\end{figure}

It is also interesting to consider these results when spectra are generated according to the specifications of existing data. In \Fig{fig:HIRES_LyaResults_Noisy} we show the same results as in the bottom panel of \Fig{fig:LyaResults} except we have adjusted the spectra to mimic HIRES spectra. Namely, we have assumed a spectral resolution with $\text{FWHM} = 6.7\kms\ $ and bins with size $\Delta v_{\text{bin}} = 2.1\kms$ (e.g. \citealt{Viel:2013fqw}). Additionally, we have assumed a signal to noise of $\text{SNR} = 10$ at the continuum per $2.1\kms\ $ pixel and that we have 20 such spectra. While we are currently only aware of 10 such spectra, this case is still interesting since spectra with significantly worse spectral resolution should also be adequate for this test.

From this figure, we can see that, despite the degradation of the spectra, the damping wing is still visible with the significance curve peaking at $\gtrsim 5\sigma$ ($\gtrsim 8\sigma$) significance for the $\axhi = 0.22$ ($0.35$) ionization state. However, this figure suggests that, in the $\axhi = 0.05$ case, it is less-clear whether the damping wing is detectable.

An important effect of adding noise to the mock spectra is that it obscures the precise location where spectra should be stacked and also increases the fraction of selected saturated regions which are, in fact, ionized. We find that for the spectra in this section $\sim 30\%$, 40\%, and 75\% of identified plausibly neutral regions are in in fact ionized for $\axhi = 0.35$, 0.22, and 0.05, respectively. This is compared to $\sim 7\%$, 10\%, and 20\% contamination when noise is neglected.

Statistical significances in this section are only estimates. In reality, the statistical significance with which the damping wing can be detected will depend on how extended the significance curves are, along with how correlated the errors in neighboring bins are. We discuss this in \S\ref{sec:Forecasts}.

\subsection{Deuterium Feature Results} \label{sec:DeuteriumResults}

We now turn to consider the prospects for identifying deuterium absorption in realistic \lyb\ mock spectra. As discussed in \S \ref{sec:RealSpectra}, our aim is to identify plausibly neutral absorption systems in the \lyb\ spectra and compare the stacked transmission moving blueward and redward away from the absorption. We identify the plausibly neutral regions in the same manner as for the damping wing.

\begin{figure}[t]
  \centering
  \includegraphics[width=9cm]{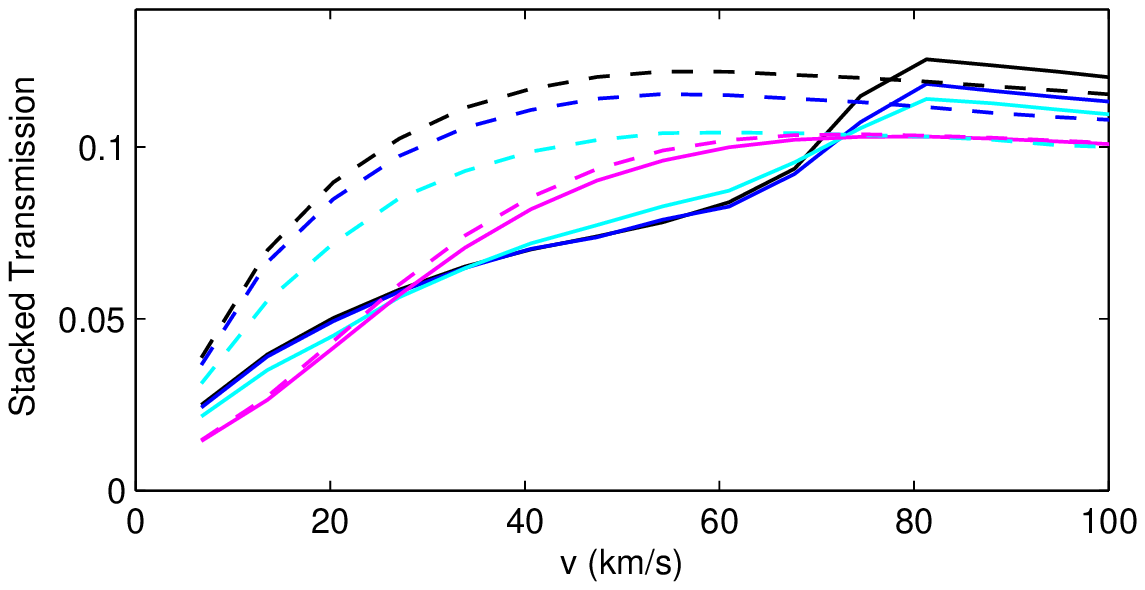}
  \includegraphics[width=9cm]{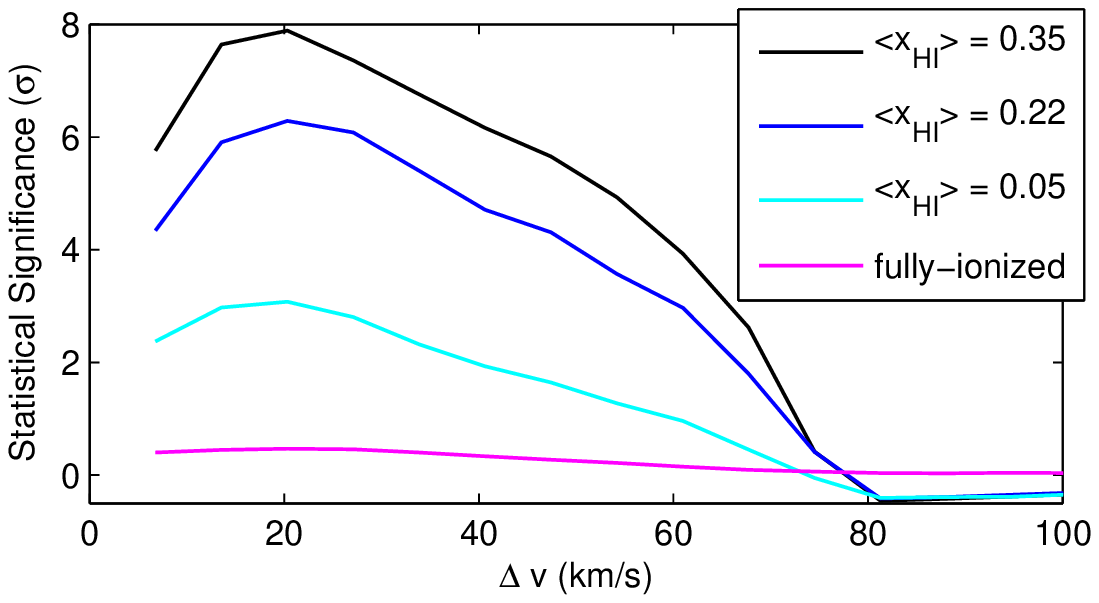}
  \caption{Deuterium \lyb\ stacking results for various neutral fractions. The top panel shows the mean ensemble-averaged noiseless stacked transmission moving blueward (solid) and redward (dashed) away from large absorption systems in the \lyb\ forest for neutral fractions $\axhi = 0.35$ (black), 0.22 (blue), 0.05 (cyan), and 0 (magenta). The bottom panel shows the excess blueward absorption in units of the standard deviation of the stacked redward transmission, assuming 20 spectra. }
  \label{fig:LybResults}
\end{figure}

In \Fig{fig:LybResults}, we show the results of applying the stacking approach to the same mock spectra as in the previous section. The top panel shows the mean transmission blueward (solid) and redward (dashed) moving away from plausibly neutral absorption systems for the same ionization states as in the previous section. We can very clearly see excess absorption in the partially neutral spectra, extending $\sim 80\kms$, consistent with our expectations from \Fig{fig:ToyD}. Additionally, we also find that, in the fully ionized case, the blueward and redward stacked transmission match up very well.

As in the previous section, we can construct a rough significance curve for the difference between the blueward and redward transmission. Specifically, in the bottom panel of \Fig{fig:LybResults} we show the excess blueward absorption in units of the standard deviation of the stacked blueward transmission assuming 20 quasar spectra. We can see that the significance of the red/blue asymmetry extends $\sim 70\kms$ ($\sim 0.3 \mpch$) and is $\gtrsim 3\sigma$ for all of the partially neutral models considered, with increasing significance for models with higher neutral fractions. Additionally, we see that the curve corresponding to the fully ionized model shows no statistically-significant deviation from red/blue symmetry. Thus, this is indeed a very clean test for the presence of deuterium. However, the signal itself is an order of magnitude smaller in velocity-space extent and is found with significantly less statistical significance than the damping wing signal. Therefore, we expect that high-resolution, high-signal-to-noise spectra will be necessary to search for it.

\begin{figure}[h]
  \centering
  \includegraphics[width=9cm]{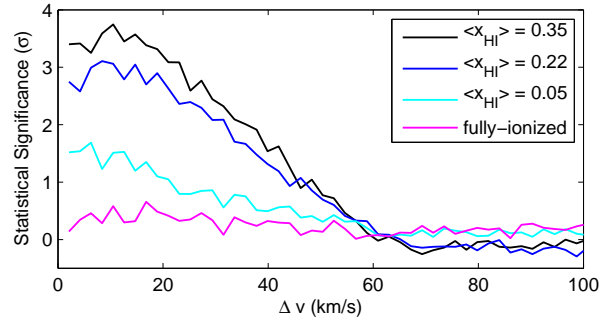}
  \caption{Results of \lyb\ stacking with HIRES-style spectra. The above panel is the same as in the bottom panel of \Fig{fig:LybResults}, except that it is generated using HIRES-style spectra, with spectral resolution of $\text{FWHM}=6.7\kms\ $ and additive noise with signal to noise of 30 per 2.1 km/s pixel at the continuum. }
  \label{fig:LybResults_Noisy}
\end{figure}

As in \S\ref{sec:Damping Wing Results}, we can reproduce \Fig{fig:LybResults} assuming spectra with specifications mimicking Keck HIRES. Unfortunately, we find that, with a signal to noise per pixel in the continuum of 10, the deuterium feature is hard to observe. Because of this, we consider using 20 HIRES-style spectra with a signal to noise per pixel of 30 in the continuum. While this signal-to-noise value is higher than those for existing spectra we found in the literature, it is not unreasonable to assume such spectra may become available in the future. Furthermore, this may provide additional motivation to obtain such spectra. Regardless, after applying the stacking approach with twenty $\text{SNR} = 30$ HIRES spectra, we obtain the results shown in \Fig{fig:LybResults_Noisy}. This figure shows that the feature should be observable with modest statistical significance. Specifically, for $\xhi = 0.35$ (0.22) the significance curve peaks at $\sim 3.7\sigma$ ($\sim 3\sigma$). Additionally, when these curves are generated assuming MIKE-style spectra, with spectral resolution $\text{FWHM} = 13.6\kms\ $ and velocity bin size $\Delta v_{\text{bin}} = 5.0\kms$, we obtain similar curves as in \Fig{fig:LybResults_Noisy} but with the signal being statistically significant over a smaller range of velocities.

Again, important effects of adding noise to the mock spectra are that it obscures the precise location where spectra should be stacked and increases the fraction of selected plausibly neutral regions which are, in fact, ionized. We find that for the spectra in this section $\sim 15\%$, 20\%, and 40\% of identified plausibly neutral regions are in in fact ionized for $\axhi = 0.35$, 0.22, and 0.05, respectively. This is compared to $\sim 7\%$, 10\%, and 20\% contamination when noise is neglected. As expected, we find a smaller level of contamination than in the previous section, owing to the increased signal to noise of the spectra used. However, for the case of deuterium, the effect of noise on the stacking location is more apparent. \Fig{fig:LybResults} demonstrates that, without noise, deuterium absorption imprints a feature on stacked noiseless spectra extending $\approx 80\kms$, but only extending $\approx 60\kms\ $ in stacked noisy spectra, as shown in \Fig{fig:LybResults_Noisy}.

The above results suggest that stacking \lyb\ transmission in high-$z$ spectra can indeed be used to detect the presence of primordial deuterium, and hence that of hydrogen, but that high-resolution and high signal-to-noise spectra will be required. Nevertheless, it would certainly be interesting to apply this approach to existing HIRES/MIKE spectra as it provides an additional test, independent of the damping wing search, for the presence of underlying neutral hydrogen in the IGM. As such, a detection with modest levels of statistical significance could lend credence to a claimed detection of the HI damping wing.

\subsection{Dark Gap Statistics} \label{sec:DarkGaps}

We now shift our focus away from stacking and toward the distribution in lengths of regions of saturated absorption (dark gaps). As discussed in \S\ref{sec:HIDistributions}, the dark gap size distribution in quasar spectra should contain information about the underlying ionization state of the IGM. Specifically, in a more neutral IGM, the typical sizes of dark gaps should be larger and the shape of the dark gap size distribution should have a more gradual decline, and possibly show a hint of bimodality, toward large $L$.

We continue this discussion in this section by considering plausible dark gap size distributions that could be observed with moderate-resolution, moderate-signal-to-noise spectra. Specifically, we consider spectra with spectral resolution $\text{FWHM} = 100\kms$, bin size $v_{\text{bin}} = 50\kms$, and a signal-to-noise ratio of 10 at the continuum. These spectra are of only slightly better quality than SDSS spectra. Additionally, since we are not concerned with \lyb\ transmission, we are able to use the entire \lya\ forest for each spectra.

\begin{figure}[h]
  \centering
  \includegraphics[width=8.5cm]{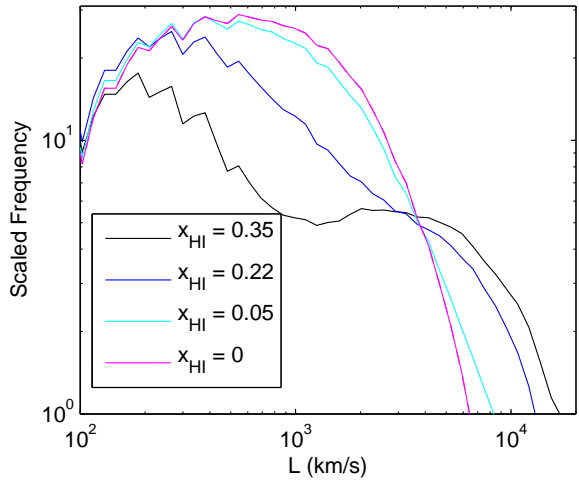}
  \caption{Mock dark gap size distribution. This figure is identical to \Fig{fig:PDFs} except that it uses spectra with spectral resolution $\text{FWHM} = 100\kms$, bin size $\Delta v_{\text{bin}} = 50\kms$, and a signal-to-noise ratio of 10 at the continuum. This figure shows the expected histogram of dark gap sizes using 20 spectra with $\axhi = 0.35$ (black), 0.22 (blue), 0.05 (cyan), and 0 (magenta) at fixed $\left\langle F \right\rangle = 0.1$.}
  \label{fig:DarkGapsResults}
\end{figure}

In \Fig{fig:DarkGapsResults}, we show the resulting dark gap size histogram expected for 20 such spectra for $\axhi = 0.35$ (black), 0.22 (blue), 0.05 (cyan), and 0 (magenta). In generating this figure, we use the same ensemble of mock spectra as in \S\ref{sec:Damping Wing Results} and \S\ref{sec:DeuteriumResults}, except with their spectral resolution and bin size modified as mentioned. We maintain the requirement that $\left\langle F \right\rangle = 0.1$.

This figure qualitatively agrees with \Fig{fig:PDFs}, where noiseless spectra with finer spectral resolution were used, but shows a shift toward larger $L$ due to smoothing the spectra. Additionally, the ionization states are not as easily distinguishable as in \Fig{fig:PDFs}, with the $\axhi = 0.05$ distribution looking practically identical to the fully ionized scenario. However, for the other neutral fractions considered, the situation looks very encouraging. The distributions for $\axhi = 0.22$ and 0.35 show a more gradual decline toward large $L$ than the fully ionized case and also reveal the clear emergence of a bimodal distribution. Additionally, the largest dark gaps in these ionization states are roughly twice as large as in the fully ionized case.

\section{Forecasts} \label{sec:Forecasts}

Having discussed the results of the proposed stacking approaches applied to realistic mock spectra, we now consider the ability of these methods to constrain the ionization state of the $z \sim 5.5$ IGM. Specifically, in this section we focus on the ability of each method to rule out the hypothesis of a fully ionized IGM.

In both the case of the HI damping wing and deuterium absorption feature, we would like to compare models representing different ionization states and estimate the $\Delta \chi^2$ between $\axhi \neq 0$ models and the fully ionized model, assuming a reasonable number of spectra. Let $F_{\axhi}(\Delta v)$ denote the mean behavior for a model with given neutral fraction, $\axhi$, as a function of stacked velocity separation and let $F_{\text{ion}}(\Delta v)$ denote the mean behavior of the ionized model, also as a function of stacked velocity separation. The precise definitions of what is meant by behavior will be discussed later in this section. In this case, the $\Delta \chi^2$ value between two models can be calculated by
\begin{align}
\Delta \chi^2_{\axhi} &= \Delta F_{\axhi} \cdot C^{-1} \cdot \Delta F_{\axhi}^{\text{T}} \label{eq:ChiSquare}
\end{align}
where $C$ is the covariance matrix of the $\axhi$ model, representing the correlation between stacked pixels, and $\Delta F_{\axhi}(\Delta v) \equiv F_{\text{ion}}(\Delta v) - F_{\axhi}(\Delta v)$, with $\Delta F_{\axhi}^{\text{T}}$ denoting its transpose. For simplicity, rather than estimating the full covariance matrix and its inverse, we approximate pixels at sufficiently wide separations as independent. We
then coarsely sample the pixels -- on the scale where they can be well approximated as independent -- and assume a diagonal covariance matrix for the coarsely sampled pixels. Specifically,
we estimate  $\Delta \chi^{2}_{\axhi}$ by simply adding up the squared statistical significance of each coarsely-sampled bin, $\left[\Delta F_{\axhi}(\Delta v_i)/\sigma_{\axhi}(\Delta v_i)\right]^2$, where $\sigma_{\axhi}(\Delta v_i)$ is the standard deviation of $F_{\axhi}(\Delta v_i)$.

\subsection{Deuterium} \label{sec:DForecast}

 Perhaps it is best to consider the case of the deuterium absorption feature first. In the case of a fully ionized IGM, the transmission looking blueward and redward from absorption systems should be symmetric on average, with excess blueward absorption only occurring when the IGM is significantly neutral. Therefore, we may calculate the $\Delta \chi_{\axhi}^2$ between stacked transmission looking redward ($F_{\axhi,\text{red}}(v)$) and blueward ($F_{\axhi,\text{blue}}(v)$) from plausibly neutral absorption systems for each ionization state $\axhi$ to estimate our ability to rule out the hypothesis of a fully ionized IGM in each case.

Thus, in the context of \Eqn{eq:ChiSquare}, we have 
\begin{align}
\Delta F_{\axhi}(\Delta v) &\equiv F_{\axhi,\text{red}}(\Delta v) - F_{\axhi,\text{blue}}(\Delta v) \\
C^{-1}_{ij} &= \delta_{ij} /\sigma^2_{\axhi,\text{blue}}(v_i)
\end{align}
where $\sigma_{\axhi,\text{blue}}(v_i)$ is the standard deviation of the stacked transmission blueward of plausibly neutral absorption systems, assuming a given number of spectra, and we have assumed that we have already resampled $\Delta F_{\axhi}(v)$ at sufficient velocity separations such that neighboring bins can be approximated as independent. At this point, the only missing ingredient is the minimum separation between two stacked pixels for them to be considered independent. We calculate the correlation function between stacked pixels in \lyb, and find that the correlation has a width of $\text{FWHM} \approx 80\kms$ and, as such, we do not expect to get more than one independent bin within the scale of the deuterium feature. Therefore, a rough estimate of the $\Delta \chi^2_{\axhi}$ value obtainable in each ionization state can be estimated simply by the peak value in the ``significance curves" in \Fig{fig:LybResults_Noisy}. Thus, if the underlying neutral fraction of the IGM is $\axhi = 0.22$ (0.35), then we expect to be able to rule out a fully ionized IGM with $\sim3\sigma$ ($\sim 3.7 \sigma$) confidence, assuming 20 HIRES/MIKE spectra with signal to noise of 30 per pixel at the continuum. Unfortunately, we do not expect to be able to rule out the hypothesis of a fully ionized IGM if the underlying neutral fraction is $\axhi \lesssim 0.05$.

\subsection{HI Damping Wing} \label{sec:HIForecast}

 Assessing the statistical significance with which we can observe the HI damping wing is slightly more complicated than the deuterium case since the test for its detection is not as simple. When faced with actual spectra, we would look for the presence of significant and extended absorption outside of large absorption systems compared to that outside of small absorption systems. 

Therefore, the behavior we would like to compare in each case is the stacked transmission outside of plausibly neutral absorption systems ($f_{\text{large}}(\Delta v)$) and the transmission outside of likely ionized absorption systems ($f_{\text{small}}(\Delta v)$). Let us denote 
\begin{align}
F(\Delta v) &\equiv f_{\text{small}}(\Delta v) - f_{\text{large}}(\Delta v)
 \end{align}
as the difference in these stacked transmissions where $F_{\axhi}(\Delta v)$ and $F_{\text{ion}}(\Delta v)$ represent this behavior for the ionization state with neutral fraction $\axhi$ and the fully ionized state, respectively. Thus, in the context of \Eqn{eq:ChiSquare}, we have
\begin{align}
\Delta F_{\axhi}(\Delta v) &= F_{\text{ion}}(\Delta v) - F_{\axhi}(\Delta v) \\
C^{-1}_{ij} &= \delta_{ij}/\sigma^2_{F_{\axhi}}(v_i) \label{eq:ChiSquare_HI}
\end{align}
 where $\sigma_{F_{\axhi}}(v_i)$ denotes the standard deviation of $F_{\axhi}(v)$ at $v_i$. The resulting $\sqrt{\Delta \chi^2}$ value indicates the expected significance with which we could rule out a \textit{fully ionized} IGM if the neutral fraction were, in fact, $\axhi$. Again, for \Eqn{eq:ChiSquare_HI}, we have assumed that $\Delta F_{\axhi}(\Delta v)$ has been resampled at velocity separations such that the pixels can be treated as independent. We find that the correlation function between pixels of stacked transmission in the \lya\ forest within the scale of the HI damping wing has $\text{FWHM}\approx 100\kms$. While this scale is large, the excess absorption due to the presence of damping wing absorption leaves a feature extending $\gtrsim 3000\kms$, leaving us with $\gtrsim 30$ independent bins within the scale of the feature.

In this manner, we are able to calculate a rough estimate for the $\Delta \chi^2$ values for the ionization states considered thus far.  Assuming the same type of spectra as in \Fig{fig:LyaResults}, namely 20 HIRES spectra with signal to noise in the continuum of 10 per pixel, we find that if the IGM is, in fact, $5\%$, $22\%$, or $35\%$ neutral, then we should be able to rule out a fully ionized IGM at $5.3\sigma$, $19.2\sigma$, or $26.3\sigma$, respectively. In the case of $\left\langle F \right\rangle = 0.05$, this reduces to $14.8\sigma$, $8.7\sigma$, and $2.2\sigma$, respectively.\footnotemark\  While we are only aware of $\sim 10$ such spectra that exist at the moment, we still regard this estimate as somewhat conservative. We found that excess stacked absorption due to the damping wing extends thousands of $\kms$, and therefore it is not necessary to have the state-of-the-art in spectral resolution to measure it. Especially with such extended correlation among neighboring pixels, it is unclear how much is gained by resolution improvements beyond $\sim 100\kms$.

\section{Conclusion} \label{sec:Conclusion}

In this work, we developed empirical tests of the possibility that the Epoch of Reionization is not yet complete by $z \sim 5.5$. Specifically, we proposed three measurements that can be made with existing, and future, high-redshift quasar spectra in order to investigate this region of reionization history parameter space.

First, we discussed using the dark gap size distribution in quasar spectra as a means of constraining the $z \sim 5.5$ neutral fraction. We find that not only do the typical sizes of dark gaps increase with neutral fraction but that the \textit{shape} of the size distribution is also sensitive to the neutral fraction. Specifically, the presence of dark gaps sourced by significantly neutral hydrogen islands introduces a bimodality in the dark gap size distribution. We find that this bimodality should be observable at $z \sim 5.5$, provided that $\axhi \gtrsim 0.2$, and should not be affected by continuum fitting errors.

Next, we proposed a method for searching for hydrogen damping wing absorption by strategically stacking regions of transmission in the \lya\ forest. Specifically, we searched for excess absorption in stacked transmission outside of plausibly neutral regions compared to that outside of likely ionized regions. We found that the presence of the hydrogen damping wing will result in excess absorption extending $\sim 5000\kms\ $ past ionization boundaries of neutral regions. Furthermore, this excess absorption should be detectable with $\gtrsim 5.3\sigma$ statistical significance for $\axhi \gtrsim 0.05$, using 20 HIRES-style spectra with a signal-to-noise value per pixel of 10 at the continuum.

Lastly, we proposed a similar stacking measurement utilizing the \lyb\ forest in order to search for deuterium absorption associated with significantly neutral hydrogen islands at $z \sim 5.5$. We proposed searching for this feature by looking for excess absorption in stacked \lyb\ transmission blueward of plausibly neutral regions compared to the corresponding redward transmission. We find that this feature should be observable in principle but will likely require additional high-resolution spectra in order to be detected. Specifically, we find that the feature should be observable at $z \sim 5.5$ with $\sim3\sigma$ ($\sim 3.7\sigma$) confidence using 20 HIRES-style spectra with a signal-to-noise value per pixel of 30 at the continuum if $\axhi = 0.22$ (0.35). While we are not aware of this many available spectra with such specifications, this provides motivation for acquiring such spectra in the future, possibly through the follow-up observation of SDSS quasars.

While we have taken many steps to ensure that the analysis of mock spectra presented in this work is realistic, there are still additional complexities that will be faced when one is presented with actual spectra. For example, we treat all portions of our spectra as being at $z = 5.5$ when, in reality, the redshift will evolve along the lines of sight. In addition, we ignored spatial fluctuations in the UV
radiation field and in the temperature density relation. Additional work will certainly be required to definitively interpret future measurements along the lines we suggest here.
However, we believe the signatures explored here are well-worth further investigation and should ultimately improve our understanding of the reionization history of the IGM.

\section*{Acknowledgements} \label{sec:ThankYou}

We would like to thank Matt McQuinn for providing us with the simulations used in this analysis. We are also grateful to Andrei Mesinger and Jordi Miralda-Escud\'e for helpful questions and discussions during the ICTP ``Cosmology with Baryons at High Redshift'' workshop.
 The authors acknowledge support from the NSF through grant AST-1109156 and NASA through grant NNX12AC97G.  

\bibliography{references_neutral_islands}

\section*{Appendix A: Contamination from DLAs?} \label{sec:DLA_contam}

A potential concern is that damping wings from super Lyman-limit systems
and damped Ly-$\alpha$ absorbers (DLAs) might produce ``false positives'' and
contaminate our search for diffuse neutral islands.
Since DLAs
and super Lyman-limit systems are mostly associated with galaxies and the
circumgalactic medium (see \citealt{Wolfe:2005jd} for a review), we would like to distinguish these absorbers 
from the more
diffuse and spatially extended islands of neutral hydrogen that are the
subject of our search. In addition, note that it is difficult to fully resolve and model high column density absorbers
in cosmological simulations (e.g. \citealt{Rahmati:2013hsa} and references therein) -- especially given our present aim of capturing the large-scale features
of reionization -- and so the impact of these systems is not captured in our present modelling.

Fortunately, we don't expect these dense absorbers to be a big contaminant,
provided we make use of the Ly-$\beta$ forest -- which helps owing to
the lower cross section in the wing of the 
line (compared to Ly-$\alpha$) -- and confine our
search to fairly extended neutral islands. The Ly-$\beta$ line profile
for a high column density absorber can be approximated
by a Lorentzian, so that the optical depth at velocity offset $\Delta v$ is:
\beqa
\tau_{\beta, \rm{DLA}} (\Delta v) = N_{\rm HI} \frac{\sigma_{\beta,0}}{\pi} \frac{R_\beta}{\left(\Delta v/c\right)^2 + R_\beta^2}.
\label{eq:taub_dla}
\eeqa
For comparison, a fully neutral and isolated absorber of co-moving length $L_{\rm neut}$ produces saturated Ly-$\beta$ absorption over a
velocity extent slightly larger than $\Delta v_{\rm neut} = H(z) L_{\rm neut}/(1+z)$. We can then
determine the column density required for a DLA to produce as long a saturated region in the Ly-$\beta$ forest as produced
by  a neutral island of
length $L_{\rm neut}$. We consider a DLA to produce saturated absorption at velocity separations where $\tau_{\beta, \rm{DLA}} \geq 3$.
Provided the extent of the absorber is large enough that $\Delta v_{\rm neut}/c \gg R_\beta$ (which
is a good approximation for the extended neutral islands of interest), this 
critical column density, $N_{\rm HI, crit}$, is given by:
\beqa
N_{\rm HI, crit} = 7.2 \times 10^{21} \rm{cm}^2 \left[\frac{\tau_{\beta, \rm{DLA}}}{3}\right] \left[\frac{1+z}{6.5}\right]
\left[\frac{L_{\rm neut}}{3.8 \rm{Mpc}/h}\right]^2. \nonumber \\
\label{eq:ncrit}
\eeqa
The fiducial value of $L_{\rm neut}$ in the above equation corresponds to $\Delta v_{\rm neut} = 500$ km/s -- this is
the minimum saturated stretch included in our stacks when we search for neutral regions (see \S \ref{sec:RealSpectra}). The 
column density $N_{\rm HI, crit}$ required for a DLA to produce this much saturated absorption is quite large, and
the abundance of DLAs with column densities larger than $N_{\rm HI, crit}$ is very small. 

Quantitatively, taking the Gamma function
fit to the column density distribution of DLAs from \citet{Prochaska:2005wy}\footnotetext{This is for the case where we do not attempt to further optimize the analysis for the decrease in transmission. It is possible that further gains could be made, with \Fig{fig:LyaResults_LowF} representing a best-possible-case scenario.} \footnote{Specifically, we use their highest redshift bin fit, 
which includes DLAs between redshifts $3.5 \leq z \leq 5.5$.} 
(which accounts for the sharp cutoff in the observed abundance of DLAs
at high column densities), we find that the number of DLAs with $N_{\rm HI} \geq N_{\rm HI, crit}$ is only
$d\mathcal{N}(> N_{\rm HI, crit})/dz = 1.5 \times 10^{-3}$. For reference, the redshift extent of the forest 
between the Ly-$\alpha$ and Ly-$\beta$
emission line at these redshifts is roughly $\Delta z \approx 1$, and so such 
high column density DLAs should be exceedingly rare. Since $N_{\rm HI, crit}$ is
only a little smaller than the exponential
cut-off in the column density distribution function, the results are rather 
sensitive to the precise choice of $N_{\rm HI, crit}$. Given that smaller column-density DLAs might still leak into our stack
if they happen to be next to saturated ionized regions, it is worth checking this dependence.
However, even choosing $N_{\rm HI,crit} = 2 \times 10^{20}$ cm$^2$ yields only $d\mathcal{N}/dz = 0.43$, which is still smaller than the abundance of neutral
islands we seek to detect. 
From these
estimates, we expect very minimal contamination from DLAs leaking into our stacked sample of possible neutral regions.
Note also that deuterium Ly-$\beta$ absorption from these high column density absorbers will be in the saturated
part of the HI Ly-$\beta$ line, and so DLAs should not contaminate our search for the deuterium signature of neutral islands either.

A separate possible worry is that DLAs could instead contaminate our sample of {\em small} saturated regions that
we use for comparison purposes (as described in \S \ref{sec:RealSpectra}). Our small saturated sample
is meant to reflect absorption around saturated yet ionized regions, and so should not contain
significant damping wing absorption. In principle, wings from any DLAs in {\em this} sample could influence the
transmission around the small saturated regions. It seems unlikely that this is a significant worry, since the
saturated yet ionized regions are likely vastly more abundant than even the small column density DLAs and super Lyman-limit systems. 
In addition, we can simply 
inspect the profile of the small saturated sample to see whether it shows any hint of damping wing absorption that
might arise from either small isolated neutral regions or DLAs.

Although contamination from DLAs does not appear to be a big worry for our tests, a more detailed examination would
certainly be warranted if possible neutral islands are discovered in real data. We may also be able to remove
DLA-contaminated regions by flagging spectral regions in the Ly-$\alpha$ and Ly-$\beta$ forest that have the same redshift
as strong metal absorbers, which generally accompany DLAs (see e.g. \citealt{Wolfe:2005jd}).

\section*{Appendix B: Further Utilizing the \lyb\ Forest} \label{sec:BetaHandle}

In order to infer the presence of the HI damping wing, we would like to compare the stacked transmission outside of plausibly neutral absorption systems to what that transmission would have been in the absence of the damping wing. Up to this point, we have been using the stacked transmission outside of \textit{small} absorption systems as a proxy for the latter quantity. From there, we argued that any extended excess absorption outside of large, plausibly neutral absorption systems compared to small, likely ionized absorption systems is indicative of the presence of damping wing absorption.

However, we do have another handle on what transmission would be like in the absence of the HI damping wing and that is the transmission in the \lyb\ forest. As discussed in \S\ref{sec:ToyD}, the HI damping wing should be significantly reduced in the \lyb\ forest compared to the \lya\ forest. Specifically, we saw that the damping wing optical depth in \lyb\ falls to less than one at velocity separations $\gtrsim 25\kms\ $ from neutral gas. Therefore, at separations greater than this, stacked transmission in the \lyb\ forest should provide information on what the shape of the \lya\ transmission would have been in the absence of the damping wing, with foreground \lya\ absorption only altering the stacked \lyb\ transmission by an overall constant $\left\langle F_{\alpha}(z_{\alpha}) \right\rangle$. Using information from stacked \lyb\ transmission has the appeal that it does not require using physically different regions of space in order to estimate the damping-wing-less transmission outside of selected plausibly neutral absorption systems. This provides protection from problems arising from unanticipated differences between the small, likely ionized absorption systems and the large, plausibly neutral absorption systems.

\begin{figure}[h]
  \centering
  \includegraphics[width=8cm]{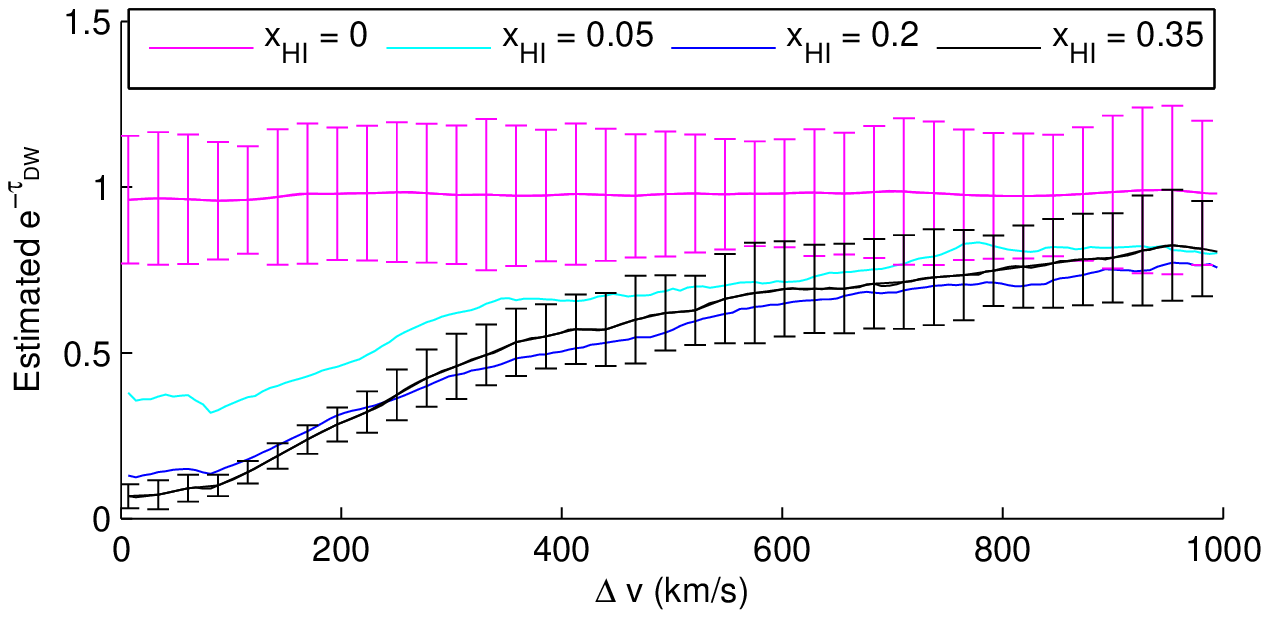}
  \includegraphics[width=8cm]{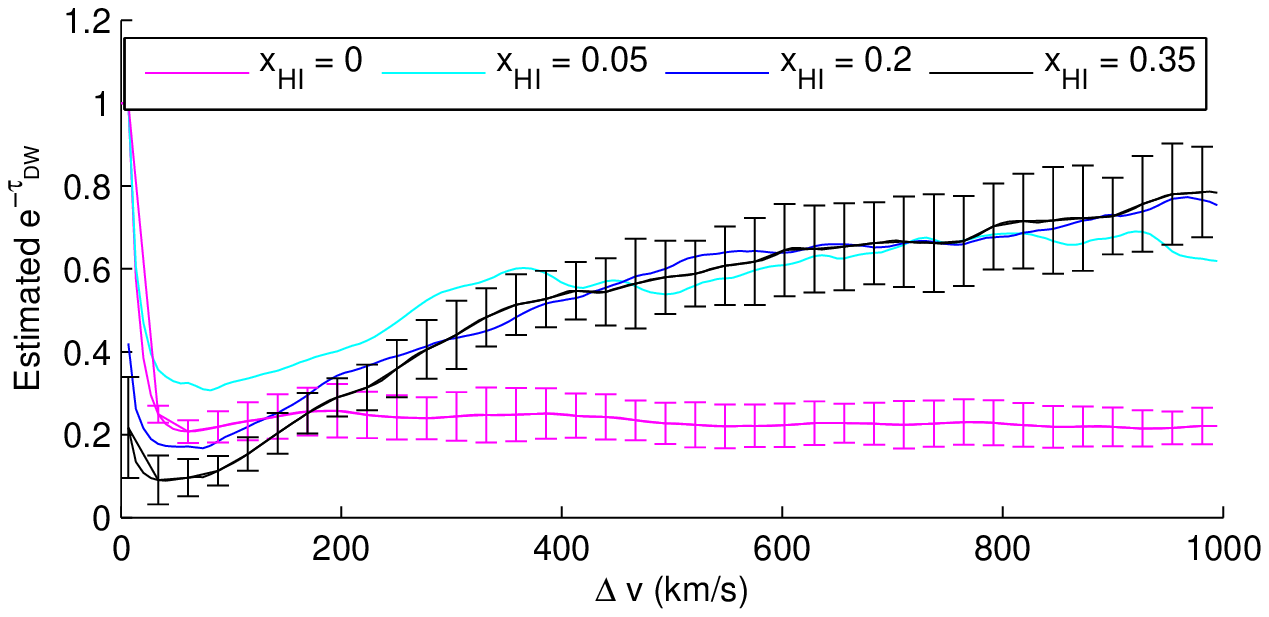}
  \caption{Using the \lyb\ forest to estimate damping-wing-less \lya\ transmission. The above figure shows the estimated \textit{shape} of stacked damping wing absorption for $\axhi = 0$ (magenta), 0.05 (cyan), 0.22 (blue), and 0.35 (black). The curves have been normalized to have their mean values peak at 1. Additionally, we show error bars for the fully ionized case and $\axhi = 0.35$ case which indicate the scatter in the curves between groups of 20 spectra. The left-hand plot is obtained by using a large ensemble of mock spectra to model a mapping between stacked \lyb\ transmission and stacked damping-wing-less \lya\ transmission and then applying this to groups of 20 spectra. Meanwhile, the right-hand figure plots the ratio of the stacked \lya\ flux to the stacked \lyb\ flux, providing a simplified estimate of the damping wing contribution to the absorption for each case. }
  \label{fig:DWShape}
\end{figure}

Thus, we would like to find a way to estimate the \lya\ transmission in the absence of the damping wing by using only the \lyb\ transmission. In principle, this can be done by using simulations to model the relationship between the two and generating a (ionization-state-dependent) mapping that takes a measurement of stacked \lyb\ transmission outside of large absorption systems and maps it to an estimate of the damping-wing-less \lya\ transmission in the same regions. From there, the ratio of the stacked \lya\ transmission to this estimate of the damping-wing-less \lya\ transmission would leave us with an estimate of $e^{-\tau_{\text{DW}}(v)}$. In the left-hand panel of \Fig{fig:DWShape}, we show the recovered $e^{-\tau_{\text{DW}}(v)}$ curve after applying this approach to each of the ionization states considered thus far, and then normalizing each curve to peak at 1. Specifically, a mapping between stacked \lyb\ transmission and stacked damping-wing-less \lya\ transmission for each ionization state was generated using a large ensemble of mock spectra and then applied to groups of 20 spectra. The error bars in the figure show the scatter in the estimated damping wing absorption between realizations of 20 spectra. For ease of viewing, we show only the error bars for $\axhi = 0$ and 0.35. This figure demonstrates that the approach works well and recovers a damping wing shape for an IGM with $\axhi \gtrsim 0.05$ (in the absence of noise).

A few things are worth pointing out about this process. First, the recovered damping wing profiles are only useful to the extent that they provide confidence that we are, in fact, observing neutral hydrogen in the IGM. The stacked profile of the HI damping wing is a complicated entity and, as such, we \textit{do not} expect to be able to, for example, fit the recovered curves to \Eqn{eq:HIDW} and estimate $\axhi$. 
Secondly, it is comforting to note that not only is no damping shape recovered in the case of $\axhi = 0$, but even if a mapping corresponding to a significantly neutral IGM is applied to a measurement of a fully ionized IGM, we do not recover a damping wing shape. Therefore, we do not expect this approach to yield false positives. Lastly, this process relies on simulations in order to map the stacked \lyb\ transmission to  damping-wing-less \lya\ transmission and is therefore somewhat model-dependent. However, we do not expect the specifics of reionization to significantly impact this mapping and we are also not interested in the fine details of the results here. We are primarily interested in whether damping wing absorption can be measured \textit{at all} in the case of a somewhat neutral IGM and, as such, we are comfortable with this level of model dependence.

Finally, we find that this mapping is relatively simple. Namely, for velocity separations $\gtrsim 100-200\kms$, the stacked \lyb\ transmission and the stacked damping-wing-less \lya\ transmission differ by roughly a constant multiplicative factor. Thus, in aiming to recover the \textit{shape} of the stacked damping wing absorption, it appears to be a very good approximation to simply divide the stacked \lya\ transmission by the stacked \lyb\ transmission. In the right-hand panel of \Fig{fig:DWShape} we simply take groups of 20 spectra and divide their stacked \lya\ transmission by their stacked \lyb\ transmission (outside of large absorption systems) and give the result unity amplitude. Qualitatively, the results look very similar to those obtained from the mapping procedure (shown on the left-hand side) but are without any model-dependence. Additionally, we again find that, in the case of a fully ionized IGM, we do not recover a damping wing shape. Thus, this provides another check which may be performed with actual spectra in order to bolster confidence that a damping wing is in fact being observed.

A potential concern for our \lya\ stacking approach in general could be that, while we make the approximation that ionized regions are exposed to a uniform ionizing background, the ionizing background will in fact be fluctuating spatially. It is then possible that, in scenarios where the ionizing background is weaker closer to the stacking location and stronger farther from the stacking location, an extended recovery could be imprinted on the stacked transmission despite the IGM being fully ionized. If one were not careful, and if these spatial fluctuations occurred on scales comparable to the damping wing feature, a false detection could be possible. One tool we have to protect against this is the fact that the scale of the damping wing is significantly smaller in \lyb\ than in \lya. Therefore, any extended recovery in stacked transmission which occurs over similar scales in \lya\ and \lyb\ is unlikely to be caused by the damping wing.

\section*{Appendix C: Extended damping wing Absorption from Correlated HI Islands} \label{sec:Correlation}

As mentioned in \S\ref{sec:Damping Wing Results}, and seen in \Fig{fig:LyaResults}, \ref{fig:LyaResults_LowF}, \& \ref{fig:HIRES_LyaResults_Noisy}, stacked \lya\ transmission outside of large saturated regions in a significantly neutral IGM displays excess absorption extending significantly past the scale of an isolated damping wing. To explain this, we seek to model the expected transmission outside of a neutral region, incorporating the correlation between the neutral region at the origin and neighboring neutral regions.  In order to simplify the calculation, while capturing the main effect, we ignore correlations between the neighbors themselves -- i.e., we only include the correlation between the neutral region at the origin and the neighbors and ignore inter-neighbor correlations.

Ignoring correlations between the neighboring neutral regions, we can approximate them as following a Poisson distribution and consider the total absorption contributed by these neutral islands or ``clouds'' following
\cite{Zuo}. Suppose that on average $m$ clouds contribute to the absorption at a given region of the spectrum.
Let $F_k \equiv e^{-\tau_1}e^{-\tau_2}\cdots e^{-\tau_k}$ denote the transmission when $k$ clouds reside along the line of sight and impact the given spectral region, with $\tau_i$ being the optical depth of the $i$th cloud. If we allow the clouds to be placed independently and if they have equal optical depths, then:
\begin{align}
\left\langle F_{k} \right\rangle &\equiv \left\langle e^{-\tau_1}e^{-\tau_2}\cdots e^{-\tau_k} \right\rangle = \left\langle e^{-\tau_1} \right\rangle \left\langle e^{-\tau_2} \right\rangle \cdots \left\langle e^{-\tau_k} \right\rangle = \left( e^{-\tau} \right)^{k}.
\end{align}
Using this expression, we can calculate the ensemble-averaged transmission by averaging over all the possible numbers of intervening clouds:
\begin{align}
\left\langle F \right\rangle &= \sum_{k = 0}^{\infty} \text{Pois}(k;m) \left\langle F_k \right\rangle = \sum_{k = 0}^{\infty} \dfrac{e^{-m}}{k!}m^{k}\left\langle e^{-\tau}\right\rangle^{k} = e^{-m}\sum_{k = 0}^{\infty} \dfrac{\left(me^{-\tau}\right)^{k}}{k!} \\
&= e^{-m}e^{me^{-\tau}} = e^{-m\left(1-e^{-\tau}\right)}.
\end{align}
Let us define the quantity $\tau_{\text{eff}}$ as
\begin{align}
e^{-m\left(1-e^{-\tau}\right)} &\equiv e^{-\tau_{\text{eff}}} \\
\tau_{\text{eff}} &\equiv m\left(1-e^{-\tau}\right).
\end{align}
We now have the absorption from neighboring clouds, characterized by the parameter $\tau_{\text{eff}•}$. 

We just need to adapt this slightly to the problem at hand. Suppose that  -- with certainty -- there is a neutral region located at $v=0$ and let us consider the excess absorption, above random, contributed
by neighboring neutral regions. 
First, the optical depth at $v = \Delta v$ that is contributed by a neighboring cloud located at $v = v'$ will depend on $|\Delta v - v'|$: clouds closer to $v = \Delta v$ will have larger optical depths at that corresponding frequency. We can account for this by substituting $\tau \to \tau(|\Delta v - v'|)$, according to \Eqn{eq:HIDW}.  Next, the expected number of HI islands in a region with velocity extent $\dd v'$ nearby our stacking location can be approximated as
\begin{align}
m &\approx \dd v' \left\langle n_{\text{HI}} \right\rangle (1 + \xi_{\text{HI,HI}}(v'))
\end{align}
where $\left\langle n_{\text{HI}} \right\rangle$ is the average number of HI islands per interval $\dd v'$ and $\xi_{\text{HI,HI}}(v')$ is the correlation function between the centers of neutral regions separated by $v'$. 
From here, we can model the effective optical depth at a given velocity separation due to neighboring HI islands, $\tau_{\text{eff}}(\Delta v)$, as
\begin{align}
\tau_{\text{eff}}(\Delta v) &= \int \dd v' \left\langle n_{\text{HI}} \right\rangle \left(1+\xi_{\text{HI,HI}}(v')\right) \left[ 1 - e^{-\tau(|\Delta v - v'|)} \right] \label{eq:taueff}.
\end{align}
In other words, the excess effective optical depth from neighboring systems involves the convolution of the absorption profile around each region with the correlation function of the neutral regions. 
This is analogous to the ``two-halo'' term in the halo model (e.g. \citealt{Cooray:2002dia}).

Thus, the model for the overall stacked transmission outside of neutral regions, also incorporating the damping wing from the central neutral region, becomes:
\begin{align}
\left\langle F \right\rangle (\Delta v) &= e^{-\tau_{\text{DW}}(\Delta v)}e^{-\tau_{\text{eff}}(\Delta v)}.
\end{align}
This model requires two inputs. First, it requires the correlation function between the centers of neutral regions, $\xi_{\text{HI,HI}}(v')$, which can be calculated from a model of the underlying ionization field. Second, the optical depth profile for neutral regions, $\tau (|\Delta v - v'|)$, is a function of the size of the neutral regions, per \Eqn{eq:HIDW}. While the neutral regions in the mock spectra take on a range of sizes, we make the simplifying approximation here -- but not in the body of this paper -- that the neutral regions at a given neutral fraction have one typical size and denote this $L_{\text{typical}}$ with a corresponding extent in velocity space $v_{\text{ext}}$. This effectively results in the optical depth profile of an individual neutral island in the right hand side of \Eqn{eq:taueff} being described by a piecewise function
\begin{align}
\tau_{\text{DW}}(\Delta v) = \begin{cases} \infty &\mbox{if } |\Delta v| < v_{\text{ext}}/2 \\ \dfrac{\tau_{\text{GP}}R_{\alpha}c}{\pi}\left[ \dfrac{1}{\Delta v - v_{\text{ext}}/2} - \dfrac{1}{\Delta v + v_{\text{ext}}/2} \right] &\mbox{otherwise.} \end{cases}
\end{align}
Next, we would like to compare this model against results using mock spectra. We do this by first generating mock spectra which \textit{only include absorption from neutral islands}, since this is the only type of absorption incorporated in our model, and stack these mock spectra at the HI/HII boundaries. To be clear, while the model curve described above adopted a fixed $L_{\text{neut}}$ for the purpose of calculating a $\tau_{\text{eff}}$, the mock spectra here are generated using the same simulated ionization fields as throughout the rest of the paper, with a wide range of sizes for the underlying neutral regions. 

To obtain a model for the stacked transmission, we first calculate the correlation function between the centers of neutral regions using the mock underlying ionization fields and also choose a value for $L_{\text{typical}}$ to be used in the optical depth profile. Additionally, in Eq. \ref{eq:taueff}, the $(1 + \xi_{\text{HI,HI}}(v'))$ term effectively breaks our integral into two pieces: the first representing the mean absorption from neutral regions and the second representing the excess or reduced absorption at $v = \Delta v$ due to the clustering of neutral islands. For our case, we are only concerned with the \textit{excess} absorption, so we make the replacement $(1 + \xi_{\text{HI,HI}}(v')) \to \xi_{\text{HI,HI}}(v')$.

\begin{figure}[h]
  \centering
	\includegraphics[width=8cm]{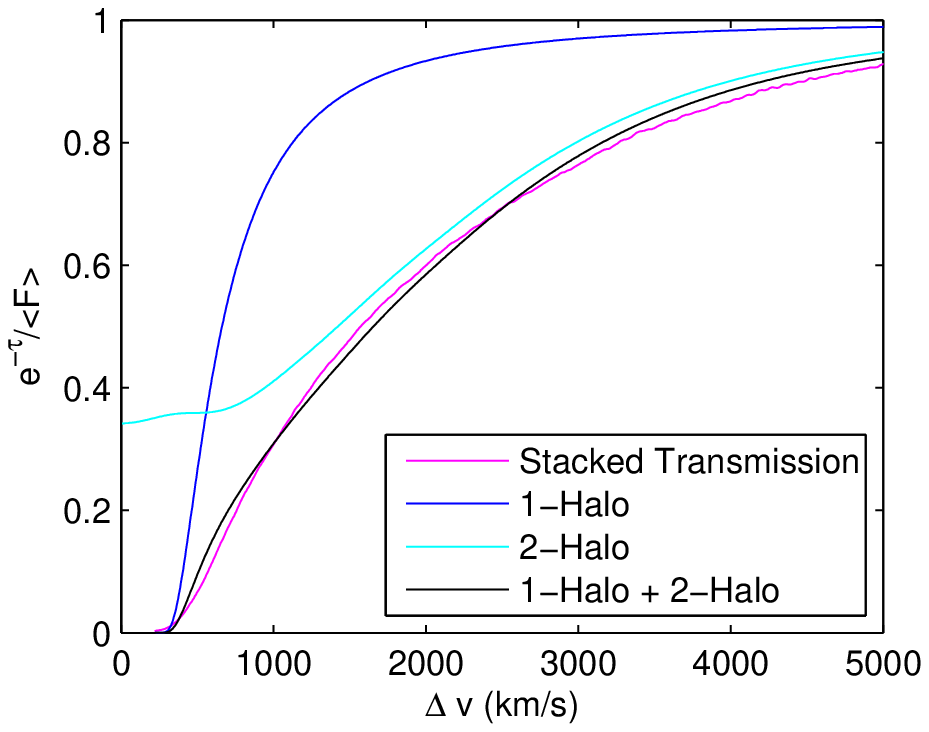}
	\includegraphics[width=8cm]{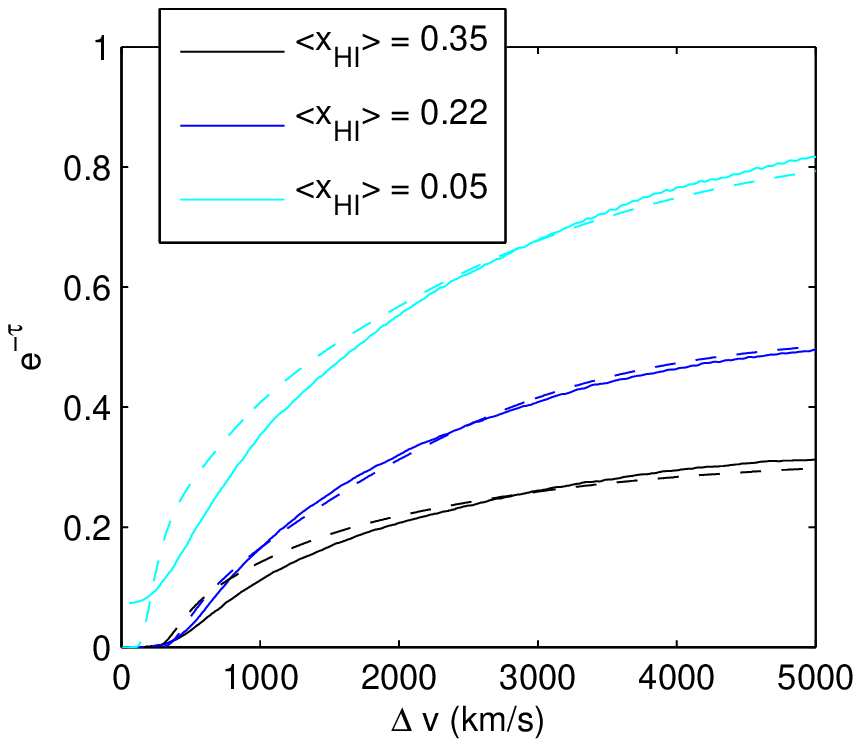}
  \caption{Model for the extended damping wing absorption. The left panel shows the components of our model for stacked transmission outside of a neutral region compared to the stacked transmission using mocked spectra (magenta) for $\axhi = 0.22$. We show the absorption due to the central neutral region (blue), average absorption due to neighboring, clustered neutral regions (cyan), and the product of the two transmissions (black). These are denoted in the legend as ``1-Halo'', ``2-Halo'', and ``1-Halo + 2-Halo'' in analogy with the halo model. In the right-hand panel, we show the comparison between the modelled transmission (dashed) and transmission from stacked mocked spectra (solid) for $\axhi = 0.35$ (black), 0.22 (blue), and 0.05 (cyan). The curves in the right-hand figure have been multiplied by the mean transmission (computed here ignoring resonant absorption for illustration).  In this appendix,  the stacking is done at the HI/HII boundaries and only damping wing absorption is incorporated to demonstrate the extended excess absorption owing to correlated neighboring systems.}
  \label{fig:HaloParts}
\end{figure}

Therefore, by providing a value for $L_{\text{typical}}$ and measuring $\xi_{\text{HI,HI}}(v')$, we can get a estimate for the mean transmission outside of neutral regions which incorporates absorption from spatially-correlated neighboring regions. In the left panel of \Fig{fig:HaloParts}, we plot an example of this for $\axhi = 0.22$. We show the modelled damping wing absorption from the central neutral region in blue, the modelled absorption from neighboring neutral islands and their damping wings in cyan, and the product of these in black. For comparison, we show the stacked transmission in the mock spectra in magenta, shifted by $v_{\text{ext}}/2$ to account for stacking occurring at HI/HII boundaries instead of at the center of neutral regions. All curves have been divided by the mean transmission. 
After taking $L_{\text{typical}} = 3.2 \mpch$, we find good agreement between the above model and the stacked transmission. The precise agreement should be taken with a grain of salt, since the model
makes several simplifying assumptions, especially that the neutral regions have a fixed size. However, the model does demonstrate that clustered neutral islands should lead to extended excess
absorption, significantly beyond the scale of an individual damping wing.

In the right panel of \Fig{fig:HaloParts}, we show the comparison between the stacked transmission (solid) and modelled transmission (dashed) for $\axhi = 0.35$ (black), 0.22 (blue), and 0.05 (cyan), where each curve has been multiplied by the mean transmission for clarity. In generating these plots, we have assumed $L_{\text{typical}} = 2.5 \mpch$, $3.2 \mpch$, and $0.75 \mpch$ for $\axhi = 0.35$, 0.22, and 0.05, respectively. We again find a very nice agreement between the modelled and stacked transmission, further confirming that spatially-correlated regions are indeed responsible for the significantly-extended excess absorption.

\end{document}